\documentclass[sigconf]{acmart}

\AtBeginDocument{%
  }

\copyrightyear{2026}
\acmYear{2026}
\setcopyright{cc}
\setcctype{by}
\acmConference[CCS '26]{Proceedings of the 2026 ACM SIGSAC Conference on Computer and Communications Security}{November 15--19, 2026}{The Hague, Netherlands}
\acmBooktitle{Proceedings of the 2026 ACM SIGSAC Conference on Computer and Communications Security (CCS '26), November 15--19, 2026, The Hague, Netherlands}
\acmDOI{10.1145/3830454.3846537}
\acmISBN{979-8-4007-2871-6/2026/11}

\usepackage{xspace}
\usepackage{fontawesome}
\usepackage{graphicx}
\usepackage{svg}
\usepackage{pifont}
\usepackage[ruled,linesnumbered]{algorithm2e}
\usepackage{multirow}
\usepackage{makecell}
\usepackage{xcolor}
\usepackage{float}
\usepackage[nameinlink,capitalize]{cleveref}

\newif\iffull
\fulltrue

\newcommand{\urlArtifacts}{\url{https://doi.org/10.5281/zenodo.22308155}}

\SetCommentSty{commentFont}

\newcommand{\confIcon}{\faLock}
\newcommand{\authIcon}{\faCertificate}
\newcommand{\intIcon}{\faShield}
\newcommand{\eaveIcon}{\faEye}
\newcommand{\mitmIcon}{\faExchange}

\newcommand{\netAttIcon}{\faSitemap}
\newcommand{\webAttIcon}{\faGlobe}
\newcommand{\affected}{\faCheck}
\newcommand{\unaffected}{~}
\newcommand{\supported}{\faCircle}
\newcommand{\partialsupport}{\faAdjust}
\newcommand{\unsupported}{\faCircleO}
\newcommand{\supportedDefault}{{\color{red}\faWarning}}
\newcommand{\fwdsupp}{\faCheck}
\newcommand{\fwdunsupp}{\faClose}
\newcommand{\fwdpublic}{{\color{red}\faShareAlt}}

\newcommand{\ssh}[1]{\texttt{#1}\xspace}

\newcommand{\sshKexInit}{\ssh{KexInit}}
\newcommand{\sshGlobalRequest}{\ssh{GlobalRequest}}

\newcommand{\sshChannelOpen}{\ssh{ChannelOpen}}
\newcommand{\sshChannelOpenConfirmation}{\ssh{ChannelOpenConfirmation}}

\newcommand{\sshChannelWindowAdjust}{\ssh{ChannelWindowAdjust}}
\newcommand{\sshChannelData}{\ssh{ChannelData}}
\newcommand{\sshChannelExtendedData}{\ssh{ChannelExtendedData}}
\newcommand{\sshChannelEof}{\ssh{ChannelEof}}
\newcommand{\sshChannelClose}{\ssh{ChannelClose}}
\newcommand{\sshChannelRequest}{\ssh{ChannelRequest}}

\begin{document}

\title{Crossing the Streams: SSH Plaintext Recovery via a Common Compression Context in Multiplexed Channels}

\author{Fabian Bäumer}
\orcid{0009-0006-5569-6625}
\affiliation{%
  \institution{Ruhr University Bochum}
  \city{Bochum}
  \country{Germany}
}
\email{fabian.baeumer@rub.de}

\author{Marcus Brinkmann}
\orcid{0000-0001-5649-6357}
\affiliation{%
  \institution{Ruhr University Bochum}
  \city{Bochum}
  \country{Germany}
}
\email{marcus.brinkmann@rub.de}

\renewcommand{\shortauthors}{Fabian Bäumer and Marcus Brinkmann}

\begin{abstract}

SSH is the standard protocol for secure remote administration of servers. At the transport layer, SSH uses the Binary Packet Protocol (BPP) for encrypted and authenticated communication. Above this, the SSH Connection Protocol multiplexes one or more logical channels over a single connection, supporting interactive shells, port forwarding, and related functionality.

We show that SSH channel multiplexing creates a previously unrecognized compression side channel: all channels on a connection share the same compression context. When compression is enabled, an attacker can inject partially chosen plaintext into a channel and observe the length of the resulting ciphertext on the network. This enables an adaptive chosen-plaintext attack that recovers secrets from one channel by interacting with another. While related attacks such as CRIME and BREACH have been studied extensively for HTTP over TLS, this is, to our knowledge, the first compression side-channel attack on SSH and the first SSH analysis to consider a combined passive eavesdropper and web attacker threat model.

We further demonstrate the attack in three different application scenarios and evaluate its effectiveness under varying levels of protocol noise. We find that, in the lowest-noise scenario, an 8-character secret over a 26-letter alphabet can be recovered using at most 276 guesses. Finally, we analyze the SSH ecosystem for compression support and other implementation characteristics that influence the practical efficacy of the attack.

\end{abstract}



\maketitle

\section{Introduction}
\label{sec:introduction}

\paragraph{Secure Shell (SSH)}

SSH~\cite{rfc4251,rfc4252,rfc4253,rfc4254} was developed as a secure replacement for Telnet and has since become the standard protocol for remote server administration, with the SSH Binary Packet Protocol (BPP) applying compression (if negotiated), padding, and authenticated encryption to messages before they are sent on the wire. However, SSH is more than just a secure terminal protocol. The SSH Connection Protocol~\cite{rfc4254} allows multiplexing of logical channels over a single encrypted connection. In addition to interactive shells, this design supports unattended commands and \emph{local} (client-to-server) or \emph{remote} (server-to-client) TCP port forwarding.

As a result, SSH is widely used as a general-purpose secure transport: users rely on SSH port forwarding to access cloud-hosted services and databases, expose local services to the Internet, and build lightweight VPN-like connections. This usage is reflected, among others, in documentation for Amazon RDS~\cite{aws-rds-ssh-tunnel}, Google Cloud services~\cite{google-ds,google-looker,google-dms}, PostgreSQL~\cite{postgresql-ssh-tunnel}, MySQL~\cite{mysql-workbench-ssh-tunnel}, public tunneling services~\cite{awesome-tunneling}, and tools such as autossh and sshtunnel~\cite{autossh,sshtunnel}. This popularity makes SSH port forwarding a security-relevant abstraction. The protocol is explicitly designed to carry heterogeneous traffic over a single connection, even if the prevalence of specific deployments is not publicly measured.

Channels of different origin and sensitivity, therefore, can share a single connection. But no part of the SSH specification considers whether this sharing has security implications: the Connection Protocol describes channels as independent streams, while the BPP compresses, pads, and encrypts the messages of all channels of a connection alike. We therefore formulate three research questions.

\begin{quote}
    \textbf{RQ1:} Can multiplexing SSH channels over a single connection create security-relevant interactions?
\end{quote}

This question is subtle because multiplexed channels are not fully independent at lower layers. Our analysis identifies transport-layer compression as a critical case: when it is enabled, data from different logical channels contribute to the same compression context before encryption, creating a side-channel leak.

Encryption of compressed plaintext can leak information through ciphertext lengths, as first described by Kelsey~\cite{FSE:Kelsey02} and later exploited in attacks such as CRIME and BREACH against HTTP over TLS~\cite{CRIME, prado2013breach}. These attacks typically require two ingredients: attacker-controlled plaintext that is compressed alongside secret data and a length oracle that reveals the size of the resulting ciphertext. In SSH, the shared compression context creates the possibility that data from different channels are compressed together, and ciphertext lengths can be observed, although with noise, by a passive eavesdropper. Turning this leakage into an attack then depends on whether the attacker can inject partially chosen data into one of those channels and filter out noise.

\begin{quote}
    \textbf{RQ2:} Can an adversary inject chosen plaintext into SSH channel data, and can channels be secret-bearing?
\end{quote}

We answer this question by analyzing SSH channel types and the application data sent over channels in \cref{sec:analysis}. We show that SSH port forwarding can provide a high-bandwidth, partially chosen-plaintext oracle within an SSH connection. Here, we consider two injection settings: a \emph{network attacker} with access to a forwarded TCP port and a \emph{web attacker} executing JavaScript in the victim's browser via a malicious site. Both can create channels but differ in oracle quality. A network attacker can inject large amounts of chosen plaintext with little noise. In contrast, a web attacker faces browser port blocking and cross-origin resource sharing (CORS) restrictions yet can trigger requests to non-blocked forwarded ports.

\begin{quote}
    \textbf{RQ3:} Is SSH channel multiplexing with compression vulnerable to an adaptive compression attack?
\end{quote}

We answer this question in the affirmative. In \cref{sec:attack}, we present a general adaptive attack strategy for SSH that combines eavesdropping with chosen-plaintext injection, assuming a combined eavesdropper and network- or web-attacker model. Building on this, in \cref{sec:evaluation}, we evaluate the attack in three proof-of-concept scenarios motivated by common uses of SSH port forwarding, measure its efficiency in terms of oracle queries, and analyze support for port forwarding and compression in SSH client implementations. Finally, in \cref{sec:x11}, we describe a server-side variant of the attack that extends its scope beyond the scenarios above.

\paragraph{Relation to Prior Attacks} 

\Cref{tab:comparison} places our work in the context of prior attacks on SSH. Existing attacks have almost exclusively targeted the Transport Layer Protocol~\cite{rfc4253}. The only prior result on the Connection Protocol~\cite{rfc4254}, the keystroke timing attack of Song et al.~\cite{USENIX:SonWagTia01}, studies a single interactive session and therefore does not cover channel multiplexing. The side channel we describe is not visible from either protocol in isolation and emerges only at their boundary, where data from independent logical channels are serialized into one stream of BPP packets with a shared compression context. Its analysis requires both protocols to be considered jointly, even though they are specified in separate RFCs and implement different abstractions. This also sets our attack apart from the compression side channels known from TLS, notably CRIME~\cite{CRIME} and BREACH~\cite{prado2013breach}, where secret and injected plaintext are carried in the same stream and the attacker writes into the compressed stream directly. In SSH, attacker-controlled bytes reach the compression context only after being wrapped in channel messages, whose headers, channel identifiers, and segmentation shape the compressed output, and the length signal is further obscured by padding and by unrelated traffic on other channels. Finally, the combined eavesdropper and web attacker model, familiar from the TLS attacks above, has not been applied to SSH before: a malicious website cannot write arbitrary bytes into an SSH channel but can only trigger browser requests that are subject to port blocking, CORS, and Private Network Access (PNA) restrictions.

\begin{table}[t]
    \centering
    \caption{Comparison of our work with prior attacks on the SSH specification. The layer column gives the targeted protocol layer, namely the Transport Layer Protocol (TLP) or Connection Protocol (CP). VSG indicates the violated security goal: confidentiality~(\textnormal{\confIcon}), authenticity~(\textnormal{\authIcon}), or integrity~(\textnormal{\intIcon}). AM lists the attacker model as either a passive eavesdropper~(\textnormal{\eaveIcon}), an active man-in-the-middle~(\textnormal{\mitmIcon}), a network attacker~(\textnormal{\netAttIcon}), or a web attacker~(\textnormal{\webAttIcon}). SCA marks side-channel attacks.}
    \label{tab:comparison}
    \begin{tabular}{llccccc}
        \toprule 
         Work & Year & Ref. & Layer & VSG & AM & SCA \\
         \midrule
         Song et al. & 2001 & \cite{USENIX:SonWagTia01} & CP & \confIcon & \eaveIcon & \faCheck \\
         Wei Dai & 2002 & \cite{weidai2002} & TLP & \confIcon & \mitmIcon & ~ \\
         Bellare et al. & 2002 & \cite{CCS:BelKohNam02} & TLP & \confIcon & \mitmIcon & ~ \\
         Albrecht et al. & 2009 & \cite{SP:AlbPatWat09} & TLP & \confIcon & \mitmIcon & \faCheck \\
         Albrecht et al. & 2016 & \cite{CCS:ADHP16} & TLP & \confIcon & \mitmIcon & \faCheck \\
         Bhargavan et al. & 2016 & \cite{NDSS:BhaLeu16} & TLP & \authIcon & \mitmIcon & ~ \\
         Bäumer et al. & 2024 & \cite{USENIX:BauBriSch24} & TLP & \intIcon & \mitmIcon & ~ \\ \midrule
         \textbf{This work} & ~ & ~ & \makecell[c]{\textbf{TLP}\\\textbf{+ CP}} & \confIcon & \makecell[c]{\eaveIcon \\ + \netAttIcon/\webAttIcon} & \faCheck \\
         \bottomrule
    \end{tabular}
\end{table}

\paragraph{Limitations} 

The scope of our results is bounded in two respects. First, we establish feasibility under concrete configurations and do not measure how often they occur in practice. Compression is almost universally supported by servers and clients but preferred by default in only four of the 33 clients we surveyed (cf. \cref{sec:evaluation:sub:implementations}); we have no evidence for the other preconditions of \cref{sec:attack:sub:model} or for their joint occurrence on one connection, so exposure remains unquantified. Second, where the preconditions do hold, exploitation is still not guaranteed: protocol noise, padding, application behavior, and browser restrictions can each degrade the length oracle or remove it altogether. We return to both points in \cref{sec:discussion}.

\paragraph{Contributions} 

This work makes the following contributions:

\begin{itemize}
    \item We perform the first security analysis of the SSH Connection Protocol, focusing on channel multiplexing and its interaction with the Binary Packet Protocol (\cref{sec:analysis}): we classify SSH channel types by whether they can carry attacker-controlled data, secret data, or both, and show that, with compression enabled, data sent on one channel can influence the compressed length of data sent on another.

    \item We present a novel adaptive compression attack that enables an adversary with eavesdropper and network attacker capabilities to recover secret data from one channel by injecting chosen plaintext in another~(\cref{sec:attack}).

    \item We evaluate the attack in three proof-of-concept scenarios: direct and browser-based injection into a forwarded TCP port, as well as recovery of an Ansible sudo password sent over a session channel~(\cref{sec:evaluation}). For the browser-based attack, we introduce a novel attacker model for SSH, the combined eavesdropper and web attacker. We also analyze a server-side attacker in a multi-user scenario~(\cref{sec:x11}).

    \item We examine the port forwarding and compression capabilities of 33~SSH clients, as well as the default settings that impact exploitability (\cref{sec:evaluation:sub:implementations}).
\end{itemize}

\paragraph{Ethical Considerations} 

Our work raises ethical concerns because the presented attacks can expose secrets of SSH users and services reached through SSH tunnels. We conducted all experiments on infrastructure under our control, collected no user data, and followed coordinated disclosure with affected vendors and the IETF sshm community. We consider publication justified because disclosure enables systematic mitigation of an existing protocol-level risk.\iffull{} For a detailed discussion of stakeholders, impacts, mitigations, and our decision to publish, we refer to \cref{sec:ethics} in the appendix.\fi

\paragraph{Open Science}

\iffull We release the paper's artifacts, including our attack implementation and raw evaluation data, under the Apache\nobreakdash-2.0 open-source license, available here: \urlArtifacts. For details, we refer to \cref{sec:openscience}.\fi

\section{Background}
\label{sec:background}

\subsection{SSH Key Exchange and User Authentication}

An SSH client connects to a server via TCP and sends a version number, a software banner, and \sshKexInit with supported algorithms, one list for each algorithm type and direction (client to server and vice versa); the server also sends its algorithm lists. Two types of algorithms are relevant to this work because they define the specifics of the transport layer of SSH, the Binary Packet Protocol~\cite{rfc4253}:

\begin{description}
    \item[Authenticated Encryption.] SSH supports common AEAD ciphers like ChaCha20-Poly1305~\cite{ietf-sshm-chacha20-poly1305-02} and AES-GCM~\cite{rfc5647, protocolopenssh}, as well as combinations of a cipher mode such as AES-CBC~\cite{rfc4253} or AES-CTR~\cite{rfc4344}, paired with a MAC such as HMAC-SHA2~\cite{rfc4253} or UMAC~\cite{miller-secsh-umac-01}.
    
    \item[Compression.] SSH compression is optional and supports only two variants based on zlib, one that compresses all messages before encryption~\cite{rfc4253} and one that delays compression until after user authentication to avoid exposing potential implementation bugs to network attackers~\cite{rfc8308, protocolopenssh}.
\end{description}

For each type and direction, the client's first algorithm supported by the server is negotiated. The client and server then perform a cryptographic handshake, which outputs a server-authenticated symmetric session key. The key is used to build a secure connection in which messages are optionally compressed and then encrypted. This transport layer is then used for user authentication by password or public key~\cite{rfc4252}. Subsequently, the client and server switch to the SSH Connection Protocol, which implements the core functionality of SSH, such as terminal sessions and network forwarding services. We describe and analyze the Connection Protocol and its interactions with the BPP in \cref{sec:analysis} of this work.

\subsection{SSH Binary Packet Protocol}

Each SSH message processed by the BPP is optionally compressed, then padded and encrypted with authentication. For \emph{compression}, SSH utilizes the zlib~\cite{rfc1950} compressed data format with the Deflate~\cite{rfc1951} compression algorithm (see \cref{sec:background:sub:compression}). The compression occurs at the BPP layer; when compression is negotiated during key exchange, each SSH message is compressed before padding and encryption. \emph{Padding} extends SSH plaintext to a multiple of the encryption block length or 8~bytes, whichever is larger. The minimum padding is 4~bytes. While SSH allows up to 255~bytes of padding to thwart traffic analysis attacks, in practice, implementations use the minimum amount of padding. The original SSH protocol only specifies \emph{encryption} algorithms that encrypt the length field. This has been used in an attack~\cite{SP:AlbPatWat09}, and as a countermeasure, new modes were proposed that do not encrypt the length field. One cipher, ChaCha20-Poly1305, preserves length field encryption while mitigating the attack by using a distinct encryption key.

\subsection{Zlib Compression}
\label{sec:background:sub:compression}

\begin{algorithm}[tb]
    \caption{LZ77 Compression (Pseudocode)}
    \label{alg:lz77}
    \SetKwInOut{Input}{Input}
    \Input{Lookahead Buffer Size $n$, Search Buffer Size $m$}
    \KwData{Input Data $D$}
    \KwResult{LZ77 Triplets $C = \{(\mathit{length}, \mathit{distance}, \mathit{next})\}$}
    $\mathit{search} \gets []$\;
    \While{$D$ is not empty}{
        $\mathit{lookahead} \gets D[0\dots n-1]$\;
        \tcc{$\mathrm{findMatch}$ finds the longest prefix of $\mathit{lookahead}$ in $\mathit{search}||\mathit{lookahead}$ which starts in $\mathit{search}$}
        $(\mathit{length}, \mathit{distance}) \gets \mathrm{findMatch}(\mathit{lookahead}, \mathit{search})$\;
        $\mathit{next} \gets D[\mathit{length}]$\;
        $\mathit{search} \gets \mathrm{lastElements}_m(\mathit{search} \| D[0\dots \mathit{length}-1])$\;
        $\mathrm{pop}(D, \mathit{length} +1)$\;
        $\mathrm{emitTriplet}((\mathit{length}, \mathit{distance}, \mathit{next}))$\;
    }
\end{algorithm}

\paragraph{LZ77} 

LZ77~\cite{lz77} is a lossless, dictionary-based compression algorithm that encodes an input stream as a sequence of \emph{(length,~distance,~next)} triplets. This approach leverages data redundancy by replacing repeated strings with back-references to their previous occurrences. In each triplet, the \emph{length} and \emph{distance} values specify a matching sequence found in the earlier data, while the \emph{next} value stores the literal immediately following that match.

For compression, the algorithm utilizes a fixed-size sliding window partitioned into a \emph{search buffer} of previously encoded data (\cref{alg:lz77}, line~1) and a \emph{lookahead buffer} of pending data (line~3). The algorithm identifies the longest prefix in the lookahead buffer that matches a sequence starting in the search buffer (line~4). If a match is found, the reference and the immediately following literal are stored (lines~4--5); if no match is found, \emph{length} and \emph{distance} are set to zero. Then, the sliding window is shifted forward by $\mathit{length} + 1$ positions (lines~6--7). The algorithm emits the resulting triplet (line~8) and repeats this process until it consumes the entire input. To decompress, the decoder sequentially copies the referenced data and appends the trailing literal for each triplet.

\paragraph{Deflate} 

Deflate~\cite{rfc1951} is a lossless hybrid compression algorithm and is extensively used in compressed data formats, for example, zlib~\cite{rfc1950} and gzip~\cite{rfc1952}. It combines LZ77 compression with Huffman coding~\cite{huffmancodes} by first applying LZ77 to the input stream---disregarding matches of length two or shorter, treating them as literals. The encoder then codes the resulting literals and matches into one of three distinct block types, selected based on which yields the smallest total footprint for a given segment of data.

Uncompressible data is handled via \emph{stored blocks}, which ignore LZ77 matches and Huffman coding and store raw input up to $2^{16}-1$ bytes. For structured data with frequent LZ77 matches, the encoder may utilize \emph{static Huffman blocks}, which employ two fixed Huffman codes defined by the standard for literal/length and distance symbols. Alternatively, \emph{dynamic Huffman blocks} improve compression for larger segments by generating optimal codes tailored to the block. While these custom codes are more efficient, they introduce additional overhead, requiring the encoder to balance the cost of the Huffman tree against the resulting bitstream savings.

\paragraph{Stream Encoding} 

The SSH specification~\cite{rfc4253} mandates the use of a \emph{partial flush} between packets to ensure no bits of the compressed SSH message in the payload remain buffered. A partial flush first closes the current block, then emits a 10-bit empty static block, which forces all data bits out as complete bytes. Due to this behavior, the number of bytes emitted during a partial flush depends on the length of the compressed {binary} packet's payload.

\section{Analysis of the SSH Connection Protocol}
\label{sec:analysis}

\subsection{SSH Connection Protocol and Channels}
\label{sec:analysis:conproto}

The central abstraction of the Connection Protocol is the \emph{channel}, which implements a typed, bidirectional byte stream with flow control. The only other abstraction is a \emph{global request} (\sshGlobalRequest), which requests connection-wide services from the peer, such as establishing or canceling a listener for remote port forwarding.

\paragraph{Channel Interface} 

\begin{figure}[t]
    \centering
    \includegraphics[width=\linewidth]{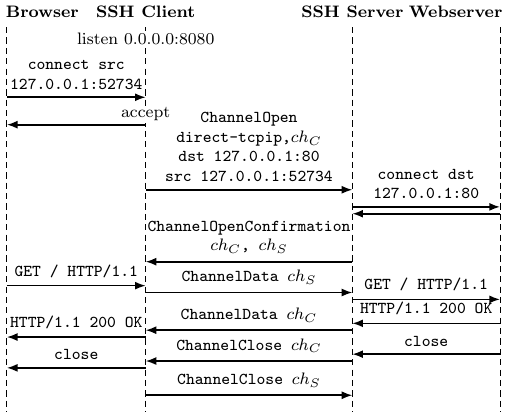}
    \caption{SSH port forwarding from local port 127.0.0.1:8080 to remote port 127.0.0.1:80.}
    \label{fig:ssh-direct-tcpip}
    \Description{A sequence diagram illustrating SSH port forwarding between a browser, an SSH client, an SSH server, and a web server. The sequence begins with the SSH client listening on 0.0.0.0:8080. The browser connects from 127.0.0.1:52734 to the SSH client, which accepts the connection and sends a direct-tcpip ChannelOpen request to the SSH server targeting destination 127.0.0.1:80. The SSH server connects to the web server and returns a ChannelOpenConfirmation. The browser then sends an HTTP GET request to the SSH client, which encapsulates it as ChannelData and sends it to the SSH server, which forwards the GET request to the web server. The web server replies with an HTTP 200 OK message, which is passed back through the SSH server as ChannelData to the SSH client and finally delivered to the browser. The process concludes with the browser and webserver closing their connections, triggering the SSH client and SSH server to exchange ChannelClose messages.}
\end{figure}

An example channel flow is shown in \cref{fig:ssh-direct-tcpip} for local port forwarding. The client prepares port forwarding by starting a TCP listener. Each time the browser connects, the client opens a new channel with the SSH server (\sshChannelOpen), which confirms it after connecting to the web server (\sshChannelOpenConfirmation). The source IP and port are provided to the server for policy enforcement but not passed through to the web server. The SSH client and server forward the TCP payload between the browser and the web server (\sshChannelData). Then the channel is closed on both sides (\sshChannelClose). This basic lifecycle interface is supplemented by messages for flow control (\sshChannelWindowAdjust, \sshChannelEof), channel-wide requests (\sshChannelRequest), and sending of typed data (\sshChannelExtendedData, used for error output in sessions).

The Connection Protocol defines \emph{channel types} to implement core SSH features, including shell sessions, command execution, forwarding of network services, X11, and the SSH agent (a secret key store for credential delegation). \cref{tab:ssh-channel-types} provides an overview of standard channel types and vendor extensions. Our use of ``client'' and ``server'' follows common use cases, although the protocol is symmetric and roles may be reversed on a per-channel basis. If the opener is listed as ``server,'' the client must first request the server to open that channel. Channels are attached to user-accessible interfaces such as pseudo-terminals, file descriptors, network ports, or UNIX domain sockets.

\paragraph{Channel Multiplexing}

\begin{figure*}
    \centering
    \iffull
    \includegraphics[width=\linewidth]{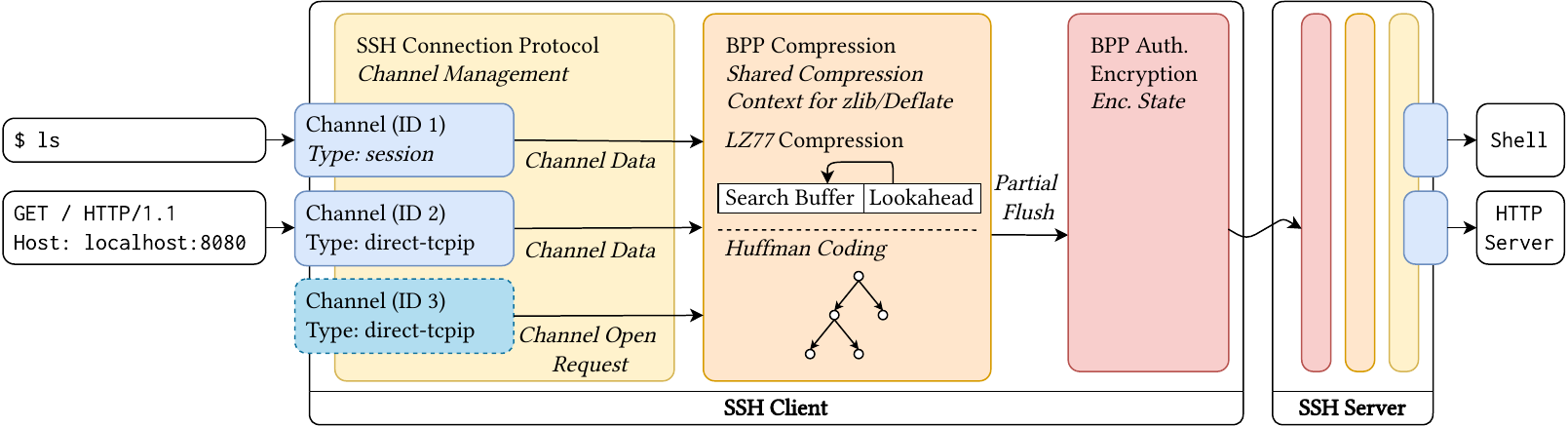}
    \fi
    \caption{An overview of SSH channel multiplexing from the client to the server. The client creates channel open requests in response to user actions or network access to a listening port for local port forwarding. Once created, each channel has a local ID, a remote ID, a type, and buffers for flow control. Data that is sent on different channels is enclosed in channel data messages, which are all compressed using a common compression context.}
    \Description{Diagram showing the architecture of an SSH session between a client on the left and a server on the right. The client side is divided into stacked layers. At the top sits the SSH Connection Protocol with channel management, holding three channels: Channel ID 1 of type session and Channel ID 2 of type direct-tcpip are already established, while Channel ID 3 of type direct-tcpip is currently being opened. Channel data from the existing channels flows down into the BPP compression layer, which contains a single shared compression context based on zlib Deflate. The compression context internally consists of a search buffer with lookahead feeding into LZ77 compression, followed by Huffman coding, and ending with a partial flush step. Below compression, the data passes through the BPP Authenticated Encryption layer, which holds the encryption state. On the server side, Channel ID 1 is connected to a shell showing a sample ls command, while Channel ID 2 is connected to an HTTP server showing a sample GET request to localhost on port 8080. The figure highlights that all channels share a single compression context but are otherwise logically separate.}
    \label{fig:ssharch}
\end{figure*}

All SSH channel messages are multiplexed over the same connection of the BPP. An overview is given in \cref{fig:ssharch}. From the user's perspective, channels are independent streams, opened and closed separately, and each regulated with its own flow-control window. But at the transport layer, all SSH messages from all channels (and any global requests) are processed sequentially by a shared BPP. If compression is enabled, each message is first compressed, then padded, encrypted, and integrity protected with the authenticated encryption cipher of the outgoing connection. On the peer side, the process is reversed: all packets are first verified and decrypted with the algorithms of the incoming connection, then, if compression is enabled, decompressed, and finally processed by the recipient channel or management layer.

\newcommand{\chName}[1]{\texttt{#1}\xspace}
\newcommand{\chc}{client\xspace}
\newcommand{\chs}{server\xspace}
\newcommand{\chC}{Client\xspace}
\newcommand{\chS}{Server\xspace}
\newcommand{\chT}{PTY/FD\xspace}
\newcommand{\chN}{TCP\xspace}
\newcommand{\chF}{UDS\xspace}

\begin{table*}[t]
  \centering
  \small
  \renewcommand{\arraystretch}{1.15}
  \caption{An overview of SSH channel types. The interface can be a pseudo-terminal or file descriptor (\chT), network port (\chN), or UNIX domain socket (\chF). \texttt{@openssh.com} abbreviated to \texttt{@}.}
  \label{tab:ssh-channel-types}
  \begin{tabular}{lccll}
    \toprule
    \textbf{Channel Type} & \textbf{Opener} & \textbf{Interface} & \textbf{Purpose} &  \textbf{Reference} \\
    \midrule
    \chName{session} & \chC & \chT & (Interactive) command execution or subsystem invocation. & \cite[Sect. 6.1]{rfc4254} \\
    \chName{direct-tcpip} & \chC & \chN & Forward data from the \chc to an outbound TCP connection in the \chs. & \cite[Sect. 7.2]{rfc4254} \\
    \chName{direct-streamlocal@} & \chC & \chF & Forward data from the \chc to a UNIX domain socket in the \chs. & \cite[Sect. 2.4]{protocolopenssh} \\       
    \chName{forwarded-tcpip} & \chS & \chN & Forward data from an incoming TCP connection in the \chs to the \chc. & \cite[Sect. 7.2]{rfc4254} \\
    \chName{forwarded-streamlocal@} & \chS & \chF & Forward data from a UNIX domain socket in the \chs to the \chc. & \cite[Sect. 2.4]{protocolopenssh} \\
    \chName{x11} & \chS & \chN or \chF & Forward authenticated X11 connections from the \chs to the \chc. & \cite[Sect. 6.3]{rfc4254} \\
    \chName{agent-connect}/\chName{auth-agent@} & \chS & \chF & Forward SSH agent connections from the \chs to the \chc. & \cite[Sect. 5.3]{ietf-sshm-ssh-agent-16} \\
    \bottomrule
  \end{tabular}
\end{table*}

\subsection{Attacker-Controlled Data and Secret Data}
\label{sec:analysis:channelinjection}

Adaptive compression attacks exploit the relative compressibility of two different types of data: secret plaintext to be recovered and attacker-chosen plaintext as candidate probes for the compressed-packet-length oracle. In this section, we examine whether SSH channel types include secret-bearing channels and whether they can provide injection vectors for a network or web attacker.

\paragraph{Channels as Injection Vectors}

To inject (partially) chosen plaintext into an SSH connection, the attacker must access the interface that the SSH client or server connects to an SSH channel. The column ``Interface'' of \cref{tab:ssh-channel-types} contains a first indicator for that ability, which we now consider in more detail.

\begin{description}
    \item[PTY/FD-connected Channels.] Interactive shell sessions connect to pseudo-terminals and remote command executions to file descriptors for standard input/output streams. In either case, access is outside our threat model, which only considers network and web attackers. We note, however, that there is a marginal risk if the attacker can control parts of the standard in- and output streams of users, for example, through log files.\footnote{A concerned SSH user raised log files as a possible injection vector for adaptive compression attacks on the OpenSSH mailing list, but the resulting discussion did not consider channel multiplexing. See \url{https://lists.mindrot.org/pipermail/openssh-unix-dev/2014-November/033176.html}, accessed 2026-04-24.}

    \item[TCP-connected Channels.] The two port forwarding types, \chName{direct-tcpip} and \chName{forwarded-tcpip}, are the primary injection surfaces for our attack, as they provide a high-bandwidth, low-noise, partially chosen-plaintext oracle. When the attacker connects to a forwarded port (or mediates such a connection through a browser), data sent to that port is split into chunks, and each chunk is prefixed by a lightweight header (message ID, channel ID, and chunk length), forming an \sshChannelData message. The additional noise on that attacker-created channel is marginal and consists of an initial \sshChannelOpen message, as well as periodical, low-frequency messages related to flow control.

    \item[UDS-connected Channels.] SSH agent and \chName{streamlocal} forwarding channels are accessed through a UNIX domain socket (UDS), that is, a local path within the filesystem namespace. However, access to such a UDS is protected by OS access controls and thus outside our threat model.

    \item[X11.] \chName{x11} channels are a special case: they can be accessed through a UNIX domain socket or a network port, depending on the implementation. Access through a UDS is protected by OS access controls and thus outside our threat model. Access to the TCP port is potentially restricted because X11 connections are authenticated. The authentication secret is a 16-byte ``MIT magic cookie,'' which is installed by the SSH server in the user's X authority database and thus outside our attacker model. However, depending on the implementation, the authentication secret may be verified at the SSH server side or at the client side. If the secret is verified at the server side, the server can block network attackers who don't know the secret from sending data over the channel. If the verification is client-side, as is the case for OpenSSH, the attacker can inject some plaintext on the channel even without knowing the secret (see \cref{sec:x11} for more details).
\end{description}
    
\paragraph{Secret-Bearing Channels} 

We now analyze each channel type for its ability to be secret-bearing, that is, to carry the same secret one or more times. For this, the interface of the channel is not a restriction because an honest application injects the secret. Instead, we ask what kind of data is transferred over the channel.

\begin{description}
    \item[Application-Defined Data Streams.] Sessions and data forwarding channel types (i.e., \chName{direct} and \chName{forwarded} channels for \chName{tcpip} and \chName{streamlocal}) are inherently candidates for secret-bearing streams, although their exploitability is situational and application-specific (see \cref{sec:evaluation:sub:ansible,sec:evaluation:sub:direct,sec:evaluation:sub:browser}).

    \item[X11.] X11 forwarding, as explained above, relies on an X11 authentication cookie. This means that each X11 forwarding channel is secret-bearing, and the secret can potentially be recovered with our attack (see \cref{sec:x11}).

    \item[SSH Agent Forwarding.] The SSH agent protocol is purposefully designed to not expose user secrets but instead to allow secure credential delegation from the client to the server. This means that these channels are not secret-bearing.
\end{description}

\paragraph{Summary} 

Out of seven SSH channel types, we have identified two that are accessible by a network attacker for partially chosen-plaintext injection and one (\chName{x11} over TCP) that may provide some restricted plaintext injection depending on implementation details, while the others are inaccessible to a network attacker because of a missing precondition (filesystem, terminal, or file descriptor access). Out of the same seven channel types, five are situationally secret-bearing because their data stream is application-defined, and one (X11) is secret-bearing by definition of the SSH Connection Protocol, while the remaining one (SSH agent forwarding) is not secret-bearing regarding private keys. We note that the two categories overlap in three cases: local TCP forwarding, remote TCP forwarding, and X11 forwarding channels can be both secret-bearing and injection vectors.

\subsection{Cross-Channel State Sharing in the BPP}
\label{sec:analysis:state-sharing}

As illustrated in \cref{fig:ssharch}, the Connection Protocol maintains separate per-channel states, such as channel identifiers, flow-control buffers, and open/closed status. However, once channel messages are handed to the BPP, they are processed sequentially within the same transport connection. We therefore analyze how data from different channels can interact through a connection-wide BPP state and what control an attacker gains through an injection vector.

\paragraph{Compression State}

If compression is enabled, SSH uses a single compression context for each direction. This context processes all messages (global requests, channel control, channel data, etc.) in the order in which they are emitted by the Connection Protocol.

In zlib-based compression, the relevant persistent state is the search buffer, which contains a $32$~KiB sliding window over previously compressed plaintext in its uncompressed form. Consequently, the compression of a message on one channel may contain LZ77 back-references to plaintext previously sent on another channel, as long as that plaintext is sufficiently close within the fully serialized message stream. Furthermore, the BPP requires a so-called ``partial flush'' at SSH message boundaries (see \cref{sec:background:sub:compression}), so each SSH message produces a complete compressed payload for one packet. While this flush drains the pending compressed output buffer and thus prevents compressed output bits from leaking across message boundaries, it leaves the search buffer intact. Message boundaries thus do not provide channel isolation, and data from different channels remain part of the same compression history until they leave the sliding window.

\paragraph{Encryption State}

The BPP encryption state is also shared across all channels in one direction: attacker-injected messages advance the packet sequence number and consume cipher output under the current keys. Correct implementations handle sequence numbers and cipher-specific usage limits by rekeying or terminating the connection before unsafe reuse occurs. This effect is distinct from the Terrapin attack by Bäumer et al.~\cite{USENIX:BauBriSch24}, which exploits sequence-number manipulation during the initial handshake rather than in the encrypted channel. In this work, we assume that SSH-au\-thenticated encryption \emph{without compression} provides the intended confidentiality and integrity guarantees, and we do not derive an attack from a shared encryption state. 

\paragraph{Summary}

A shared BPP state has different security implications for compression and encryption. For encryption, we rely on the security of the underlying authenticated encryption scheme. For compression, however, the shared search buffer creates a direct cross-channel dependency: attacker-controlled data on one channel and secret-bearing data on another channel can affect each other's compressed length within the same compression window. This cross-channel dependency is the root cause exploited by our attack.

\section{Adaptive Compression Attack on SSH}
\label{sec:attack}

\subsection{Attacker Model}
\label{sec:attack:sub:model}

We separate the attacker's capabilities from the preconditions that must hold in a vulnerable deployment.

\paragraph{Capabilities} 

The attacker is passive regarding the SSH connection and can observe the total length of ciphertexts transmitted during a time interval, either by eavesdropping on the network path or through another length side channel (see \cref{sec:x11}). Additionally, the attacker can inject partially chosen plaintext into the SSH connection, either as a network or web attacker:

\begin{itemize}
    \item A \emph{network attacker} connects directly to a TCP port that the victim's SSH client or server has exposed through forwarding, which provides a high-bandwidth, low-noise oracle.
    \item A \emph{web attacker} controls a website that the victim visits while port forwarding is active; JavaScript on the site makes the victim's browser issue HTTP requests to a forwarded port.
\end{itemize}
We discuss both settings in detail in \cref{sec:evaluation:sub:direct,sec:evaluation:sub:browser}.

\paragraph{Preconditions} 

For the attacked SSH deployments, we assume that the victim has enabled SSH compression for the vulnerable connection. We also assume that several channels are multiplexed over the same connection in the same direction, including at least one secret-bearing channel and one injection channel (described above). Finally, we assume that the secret is transmitted multiple times, either within the same connection or across multiple connections during the same attack.

 All capabilities and preconditions must hold simultaneously for the attack to succeed; the assumptions below affect only its cost.
 
\paragraph{Cost-Easing Assumptions} 

In our evaluation, we make three more assumptions that reduce the cost of the attack. We assume that all secrets are prefixed by a known plaintext string of at least two bytes to anchor the attacker's secret guess; otherwise, the attacker has to brute-force guess the leading bytes of the secret, which can then serve as the anchor for the remaining bytes (see \cref{sec:attack:sub:algorithm}). We also assume that the attacker knows the alignment length that lands a guess on a tipping point so that the length difference between a correct and an incorrect candidate survives the SSH padding; otherwise, the attacker has to sweep all alignment lengths in every measurement, which multiplies the number of measurements per round and requires a higher commit margin (see \cref{sec:attack:sub:optimization}). Furthermore, we assume that the attacker can trigger the secret's transmission instead of waiting for the transmission and detecting it through another side channel. If any of these cost-easing assumptions do not hold, the attack can still succeed at a higher cost. We discuss the feasibility of these conditions in \cref{sec:discussion:sub:feasability}.

\paragraph{Unconstrained Parameters} 

Our model does not constrain the negotiated authenticated encryption mode, nor does it require the attacker to identify the boundaries between individual binary packets. In particular, the attack can be applied even when the cipher encrypts the binary packet's length field. The attacker only needs to observe the aggregate ciphertext length over the measurement interval, rather than the length of any individual SSH packet.

\subsection{Attack Algorithm}

Given the attacker model of \cref{sec:attack:sub:model}, the algorithm recovers the secret one byte at a time, similar in spirit to CRIME~\cite{CRIME} and BREACH~\cite{prado2013breach}. \cref{alg:attack} gives the pseudocode.

\label{sec:attack:sub:algorithm}

\begin{algorithm}[tb]
    \caption{Adaptive Compression Attack (Pseudocode)}\label{alg:attack}
    \SetKwInOut{Input}{Input}
    \Input{Known prefix $p$, alphabet $A$, alignment length $\ell$, commit margin $\mu$}
    \KwResult{Recovered secret $s$}
    $s \gets \varepsilon$\tcp*{empty string}
    \Repeat{$\mathrm{secretIsFullyRecovered}(s)$}{
    $\mathit{sum}[c] \gets 0$ for each $c \in A$\;
    \Repeat{$\mathit{sum}[b'] - \mathit{sum}[b] \ge \mu$}{
    \ForEach{$c \in A$}{
    $\mathrm{flushLZ77SearchBuffer}()$\;
    $\mathrm{triggerOrWaitForSecret}()$\;
    $\mathrm{injectPlaintext}(p \| s \| c \| \mathrm{align}(\ell))$\;
    $\mathit{sum}[c] \gets \mathit{sum}[c] + \mathrm{measureTransferredData}()$\;
    }
    $b \gets c \in A$ with smallest $\mathit{sum}[c]$\tcp*{best candidate}
    $b' \gets c \in A \setminus \{b\}$ with smallest $\mathit{sum}[c]$\tcp*{2nd best}
    }
    $s \gets s \| b$\;
    } 
\end{algorithm}

\paragraph{Leaking a Single Byte}

The inner loop of \cref{alg:attack}~(lines~5--10) issues one trial per candidate byte. Each trial flushes the LZ77 search buffer with $32$~KiB of random data~(line~6), triggers or waits for a transmission of the secret~(line~7), injects the guess $p \mathbin{\|} s \mathbin{\|} c \mathbin{\|} \mathrm{align}(\ell)$~(line~8), and measures the resulting wire-transmission size for a fixed amount of time immediately following the guess's injection~(line~9). Measuring for a fixed amount of time solves two problems: first, the attacker cannot reliably determine at which exact time the guess is sent over the wire; second, when the binary packet's length field is encrypted, packet boundaries may be opaque to the attacker. The measurement's time window should be as short as possible, allowing the encrypted guess to be included but subsequent transmissions to be excluded. Here $p$ is a known prefix preceding the secret, $s$ are the recovered secret bytes, $c$ is the candidate byte under test, and $\mathrm{align}(\ell)$ is the alignment of length~$\ell$.

The guess is structured in such a way that the compressed length is likely to reveal whether $c$ is correct. When $c$ is correct, $p \mathbin{\|} s \mathbin{\|} c$ matches a substring already present in the search buffer from the recent secret transmission, and LZ77 replaces it with a back-reference; an incorrect $c$ breaks the match and forces the byte to appear as a literal, thereby increasing the length of the compressed plaintext. Deflate's minimum match length is three, so $p$ and $s$ together must contribute at least two bytes; a longer prefix also reduces the chance of an undesired match against unrelated content. Without a known prefix, the attacker first brute-forces the leading bytes, and the recovered material then serves as the prefix.

The alignment $\mathrm{align}(\ell)$ pads the guess to a \emph{tipping point}: the length at which the extra literal of a wrong candidate forces the BPP encryption to emit one additional ciphertext block, amplifying a few-bit difference in the compressed plaintext length into a multiple-byte length difference observable on the wire. Alignment bytes must therefore avoid further LZ77 matches of their own. We additionally restrict them to bytes that Deflate's static Huffman code emits as 8-bit literals so that each alignment byte advances the encoded length by one byte and every tipping point is reachable.

\paragraph{Flushing the LZ77 Search Buffer}

After each guess, the search buffer contains the previous guess and any noise the oracle introduced. A fresh secret transmission alone does not guarantee that the next round matches the secret rather than a previous guess: Deflate's $\mathrm{findMatch}$ is a hash-chain search and does not always return the closest match. Injecting $32$~KiB of random data, the maximum search buffer size in Deflate, evicts every prior byte and reliably eliminates these residual matches, although it may introduce some channel data headers along with it due to segmentation.

\paragraph{Committing a Byte}

A single measurement of the transmission size is noisy. The algorithm therefore repeats the measurement and sums the per-trial sizes across rounds~(line~9); the noise mostly averages out in the sum, while the gap between the correct candidate and the others keeps growing. Rather than fixing the number of rounds in advance, the algorithm stops repeating as soon as one candidate stands out from the others. Let $b$ be the candidate with the smallest sum and $b'$ the second-smallest~(lines~11--12). The algorithm commits $b$ once $\mathit{sum}[b'] - \mathit{sum}[b] \ge \mu$~(line~13). We call $\mu$ the \emph{commit margin}. A larger $\mu$ takes longer and thus requires more secret transmissions but is less likely to commit a wrong byte.

\paragraph{Full Secret Recovery}

The outer loop~(lines~2--3 and 14--15) extends a single committed byte to the full secret. Once a byte is committed, the algorithm appends it to $s$, and the recovered material then serves as the prefix for the next round of inner-loop measurements. The algorithm needs a termination condition: if the secret ends with a known terminator, for example, a zero byte or a line feed, the attack stops once the terminator is committed. If the secret carries an explicit length prefix, the algorithm recovers that length first and uses it to bound the loop. For passwords, an authentication attempt may serve as a check instead.

\subsection{Noise Compensation Strategies}
\label{sec:attack:sub:optimization}

The basic algorithm of \cref{sec:attack:sub:algorithm} reliably recovers a secret from an encrypted and compressed SSH connection, but practical exploitation is still costly. Each measurement requires another transmission of the secret, and noisy environments require many repetitions to push the signal above the noise. The algorithm also assumes the attacker knows the correct alignment length $\ell$ in advance---the value that lands the guess on a tipping point. Below, we describe three generic compensation strategies that reduce the impact of noise and let the attack proceed with an unknown alignment length.

\paragraph{Candidate Elimination}

In the basic algorithm, a loop~(lines~5--10 of \cref{alg:attack}) measures every candidate in every round until one commits. This is wasteful: once a candidate has accumulated a much larger $\mathit{sum}[c]$ than the leader, it is unlikely to overtake it, yet the algorithm keeps measuring it every round. We instead eliminate such candidates early. We could choose the elimination threshold independently, but we reuse the commit margin $\mu$: whenever $\mathit{sum}[c] - \mathit{sum}[b] \ge \mu$ for the best candidate $b$, the algorithm drops $c$ from subsequent rounds for the current byte.

\paragraph{Full Alignment Sweep}

To avoid the requirement of knowing $\ell$ in advance, the attacker can sweep over all possible alignment lengths in every measurement: try each $\ell$ once per candidate and accumulate $\mathit{sum}[c]$ across all alignments. For a wrong alignment, the SSH padding swallows the per-candidate compression difference, so every candidate contributes the same number of bytes to $\mathit{sum}[c]$. Only the correct alignment contributes a per-candidate signal, and the sum across all alignments still reflects it. The price is more measurements per round and amplified noise, which can require a noticeably higher commit margin.

\paragraph{Adaptive Alignment Sweep}

A modified variant of the previous compensation strategy prunes unproductive alignment lengths to reduce the number of measurements per round. An alignment length is unproductive when measurements show no observable difference across candidates for that alignment: the SSH padding has absorbed the signal, indicating that the alignment is not at a tipping point. Pruning is optimistic---with non-negligible probability, the correct alignment looks unproductive in early rounds, and the algorithm prunes it anyway. To recover, the algorithm reintroduces pruned alignment lengths when the remaining ones produce no observable difference for two consecutive rounds. The pruned set also carries over between byte positions.

\subsection{Differences from TLS Compression Attacks}
\label{sec:attack:sub:tls-diff}

Although our attack builds on the same general idea as adaptive compression attacks on TLS, such as CRIME~\cite{CRIME} and BREACH~\cite{prado2013breach}, differences in the protocol require significant adjustments. SSH ciphers such as ChaCha20-Poly1305 encrypt the BPP length field, so the attacker cannot reliably read individual packet sizes; CRIME's per-record length measurements do not apply, and the attacker has to recover the per-byte signal from the number of transmitted bytes. The attacker also cannot reset the SSH compression state mid-connection: where CRIME and BREACH rely on a fresh HTTP request to cycle prior content out of the compression window, the SSH attacker has to actively flush $32$~KiB of random data before every measurement, which inflates the attack's bandwidth cost. Furthermore, the attacker has no direct write access to the compression input---when the attacker writes to a forwarded TCP port, the SSH client wraps those bytes in channel-data messages, and the framing headers add their own implementation-dependent noise.

These differences are not incidental: the injection oracle itself is different from those in TLS. SSH multiplexes logically separate channels over a single compressed connection, so the secret-bearing and the attacker's injection channel can share a compression context even when independent at the application layer. This cross-channel compression context has no analog in TLS, and to our knowledge, this is the first adaptive compression attack to exploit it.

\section{Evaluation}
\label{sec:evaluation}

In this section, we evaluate the generic adaptive compression attack of \cref{sec:attack} in three proof-of-concept scenarios. Each scenario instantiates a partially chosen-plaintext oracle of the kind discussed in \cref{sec:analysis:channelinjection} but combines a different set of SSH features and attacker capabilities: direct injection through a network-exposed forwarded port (\cref{sec:evaluation:sub:direct}); browser-based injection through the victim's own browser into a loopback-bound forward (\cref{sec:evaluation:sub:browser}); and recovery of an Ansible privilege-escalation password transmitted on a session channel, with plaintext injection through a forward inherited from the user's SSH configuration (\cref{sec:evaluation:sub:ansible}). \cref{sec:evaluation:sub:performance} compares the practical performance of these variants across different configurations, and \cref{sec:evaluation:sub:implementations} surveys SSH implementations that meet the preconditions for the attack.

We target OpenSSH, one of the most widely deployed SSH implementations, and negotiate ChaCha20-Poly1305 as the authenticated-encryption mode---the default in recent OpenSSH releases. Because this mode encrypts the length field of every SSH binary packet, our results prove that the attack remains effective even when the length field of the BPP is encrypted and the individual size of a packet cannot be reliably determined by an eavesdropper.

For all variants, our proof-of-concept implementation grants the attacker the ability to actively trigger the transmission of the secret, reducing runtime and eliminating timing variability that would otherwise complicate reproducibility. This is only a shortcut; an attacker who has to wait for the secret can mount the same attack, just more slowly; see \cref{sec:attack:sub:model}. The measurement of a guess's transmission size is done within a fixed amount of time immediately following the guess's injection; the amount of time differs between our three scenarios to account for scenario-specific delays introduced through the injection oracle.

\subsection{Direct Plaintext Injection}
\label{sec:evaluation:sub:direct}

\begin{figure}[tb]
    \centering
    \iffull
    \includegraphics[width=\linewidth]{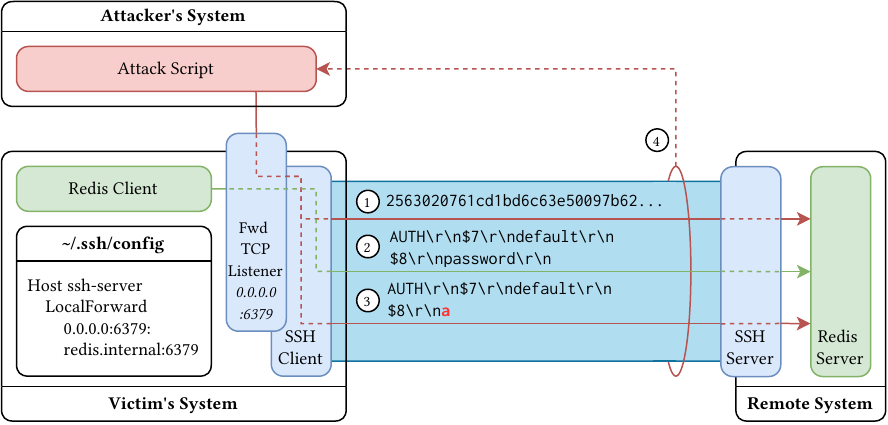}
    \fi
    \caption{Attack flow for a single guess using direct injection (cf.\ \cref{sec:evaluation:sub:direct}). \ding{172}~First, the attacker flushes the compression state by transmitting 32~KiB of random data, evicting any prior secret or guess from LZ77's search buffer. \ding{173}~Next, the attacker triggers or waits for the transmission of the secret before \ding{174}~injecting their guess. \ding{175}~Finally, the attacker measures the transmitted data size and compares it to the measurements of the other candidates.}
    \Description{A system diagram shows a victim's system with a Redis client and an SSH client. The configuration file indicates a local port is forwarded to a remote Redis server. Three data streams pass through the SSH tunnel. The first is a long string of random characters. The second is a Redis authentication message including a full password. The third is a Redis authentication message containing only one character of a password guess. An attacker is depicted monitoring the encrypted connection between the SSH client and the SSH server.}
    \label{fig:variant-direct}
\end{figure}

In the first variant, we target the password of a Redis authentication request that the victim sends over a locally forwarded port. We consider a network attacker who can connect to that forwarded port directly; on OpenSSH, this requires the victim to either specify an explicit bind address or enable the \texttt{GatewayPorts} option.

Direct access to the forwarded port lets the attacker open arbitrary TCP connections and, therefore, arbitrary logical channels within the SSH connection. Any data sent over such a channel enters the client-to-server compression context, providing a partially chosen-plaintext oracle. \cref{fig:variant-direct} illustrates the procedure for a single guess. First, the attacker evicts any prior secret or guess from the LZ77 search buffer by transmitting $32$~KiB of random data. This step is necessary because the SSH session persists across rounds; without it, residual data from earlier iterations would remain in the buffer. The attacker then triggers or awaits the next transmission of the secret and immediately injects a guess through the oracle. Finally, the size of the resulting transmission is measured and compared against the measurements of the other candidates, following the generic algorithm of \cref{sec:attack:sub:algorithm}.

\subsection{Browser-Based Plaintext Injection}
\label{sec:evaluation:sub:browser}

\begin{figure}[tb]
    \centering
    \iffull
    \includegraphics[width=\linewidth]{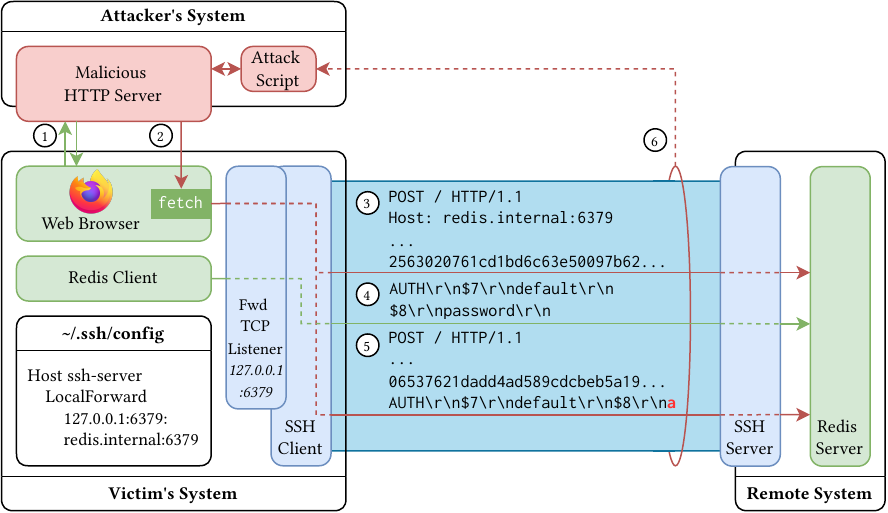}
    \fi
    \caption{Attack flow for a single guess using browser-based injection (cf.\ \cref{sec:evaluation:sub:browser}). \ding{172}~First, the victim visits an attacker-controlled website, and \ding{173}~the attacker establishes a WebSocket-based control channel for the injection of guesses. \ding{174}~As in direct injection, the compression state is then flushed with a sufficiently long request. \ding{175}~Next, the attacker triggers or waits for the transmission of the secret, \ding{176}~injecting their guess immediately after. To account for dynamic Huffman coding, the attacker prepends random data to the request body, isolating the guess in a separate static Huffman block. \ding{177}~Finally, the attacker measures the transmitted data size and compares it to the measurements of the other candidates.}
    \label{fig:variant-browser}
    \Description{A system diagram shows a victim's system communicating with remote servers through an SSH tunnel. Inside the victim's system are a web browser, a Redis client, and an SSH client. An SSH configuration file defines a local port forward to a remote Redis server. Outside the system, an attacker controls a malicious HTTP server that sends requests to the web browser. Three numbered data streams flow through the SSH tunnel. The first stream shows an HTTP POST request containing a long hexadecimal string. The second stream shows a Redis authentication message with a full password. The third stream shows an HTTP POST request combined with a partial Redis authentication message. An attacker is shown monitoring the encrypted connection between the SSH client and the SSH server.}
\end{figure}

For the second variant, we remove the attacker's ability to connect to the forwarded port directly by assuming the port is bound to \texttt{localhost}, the OpenSSH default. As a consequence, direct TCP injection is no longer possible. Instead, we consider the combined passive eavesdropper and web attacker model in which the victim visits an attacker-controlled website that injects guesses through JavaScript. Because the browser runs on the same host as the SSH client, it can reach the loopback-bound forward.

Aside from an initial website visit and the establishment of a WebSocket-based control channel, the attack steps remain largely unchanged compared to direct injection; \cref{fig:variant-browser} illustrates a single guess round. Each injection now constitutes an HTTP request, whose surrounding protocol metadata introduces measurement noise that we average out across rounds. The HTTP framing additionally biases the Deflate encoder toward dynamic Huffman blocks, in which alignment bytes no longer encode to the uniform 8-bit codes the alignment sweep relies on. The attacker counters this by prepending random data to the guess body, isolating the guess in a small static Huffman block where the per-byte signal is restored.

The proof of concept employs a Playwright-automated Firefox to simulate the victim's browser. We choose Firefox specifically because it does not implement Private Network Access (PNA), a cross-site request forgery defense that Chromium- and WebKit-based browsers apply to local services. This lets us reuse the same Redis target as in the direct variant so that the two variants differ only in their injection mechanism and can be compared directly.

\paragraph{Remark: Restricted Plaintext Injection Oracle in CORS-PNA}

The attack remains feasible, with some limitations, even when PNA is enforced. A browser with PNA preempts the cross-origin injection with an \texttt{OPTIONS} preflight that strips the request body, but the attacker can still hide a guess in the preflight's URL path or a header name if the secret is restricted to a small alphabet: the letters \texttt{A}--\texttt{Z} and \texttt{a}--\texttt{z}, the digits \texttt{0}--\texttt{9}, and a fixed set of printable punctuation marks, the composition of which is field-specific. Crucially, the alphabet excludes whitespace and all control bytes and, hence, the carriage return and line feed that terminate every line of the Redis wire format. As a consequence, at least three password characters (the minimum length of a backreference) must be recovered by brute force before the attack can proceed normally (for targets besides Redis, brute force is not required if a long enough prefix can be represented in the restricted PNA alphabet). The limited alphabet also reduces the effectiveness of the flush by increasing the likelihood of dynamic Huffman codes and backreferences against the flush. From version~142 onward, Chromium additionally requires a one-time user permission for local-service connections. We evaluated Chromium~149 under the browser-based margins from \cref{tab:evaluation} and found a 67--69\% success rate (out of 100 trials) with guess counts ranging from $2\,867$ to $40\,326$ per trial in a known-prefix scenario.

\subsection{Ansible Password Recovery}
\label{sec:evaluation:sub:ansible}

\begin{figure}[tb]
    \centering
    \iffull
    \includegraphics[width=\linewidth]{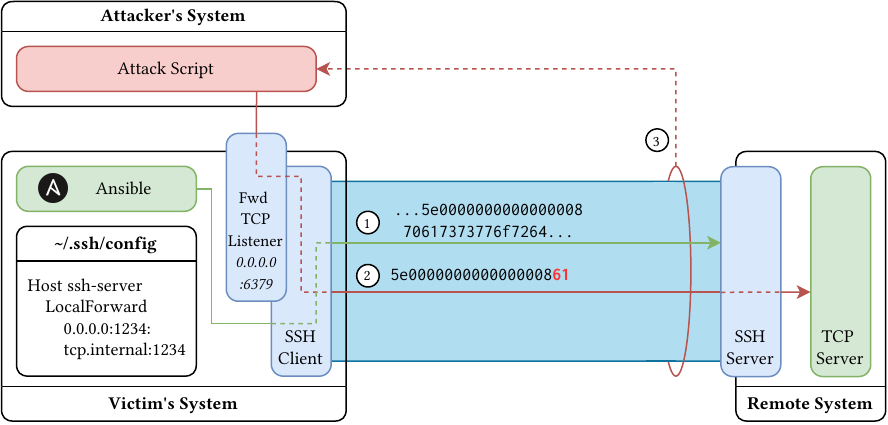}
    \fi
    \caption{Attack flow for a single guess against an Ansible \texttt{become} password (cf.\ \cref{sec:evaluation:sub:ansible}). As in direct injection, \ding{172}~the attacker triggers or waits for the transmission of the secret before \ding{173}~injecting their guess. \ding{174}~Finally, the attacker measures the transmitted data size and compares it to the measurements of the other candidates. Compared to direct injection, no attacker-initiated flush of the compression context is required since the secret is transmitted at the beginning of a fresh SSH connection.}
    \label{fig:variant-ansible}
    \Description{A system diagram shows a victim's system containing an Ansible playbook and an SSH client. The SSH client is configured with a local port forward to a remote TCP server. Two numbered data streams pass through the SSH tunnel. The first stream contains a long hexadecimal string that includes the encoded word ``password''. The second stream contains a shorter hexadecimal string ending in the character ``a''. An attacker is shown monitoring the encrypted traffic between the SSH client and the SSH server.}
\end{figure}

For the third variant, we extend our attack to a secret carried on an SSH channel type besides \chName{direct-tcpip}. Ansible is an open-source infrastructure automation tool that runs predefined sets of tasks, called playbooks, against remote hosts. We target the password used by Ansible's \texttt{become} plugin, which elevates individual tasks to another user via \texttt{sudo}. The \texttt{become} plugin writes the user's password to the SSH client's \texttt{stdin}, where it is transmitted within a \chName{session} channel alongside a predictable channel-data header. Knowing this prefix, we can apply the same byte-by-byte recovery as in the direct variant; \cref{fig:variant-ansible} illustrates a single guess. Compared to \cref{fig:variant-direct}, no flush step is required: each playbook invocation establishes a fresh SSH connection with an empty compression context, eliminating prior-state eviction and minimizing measurement noise.

Three aspects of Ansible's default behavior are particularly relevant. First, Ansible enables SSH compression by default, so every Ansible-driven SSH connection is susceptible to compression-based side channels. Second, Ansible invokes the local system's \texttt{ssh} binary without a dedicated configuration file, loading the user's defaults instead; any \texttt{LocalForward} directive the user configured at some point is therefore inherited by playbook executions and may serve as an injection vector. Third, Ansible reuses a single SSH master connection across tasks via the \texttt{ControlMaster} feature, with a default timeout of 60~seconds. We disabled this feature in our proof of concept to simulate playbook runs more than 60~seconds apart, so each invocation creates a fresh connection. The attack remains applicable when \texttt{ControlMaster} is active by reintroducing the flush from the direct variant; since the master itself is established with compression by default, every task multiplexed over it shares the same compression context and is vulnerable to the same attack.

\subsection{Attack Performance}
\label{sec:evaluation:sub:performance}

We evaluate the practical performance of our three client-side attack variants in our proof-of-concept implementation. For each variant, we attempt to recover an 8-character lowercase password under five configurations spanning the compensation strategies of \cref{sec:attack:sub:optimization}: no further compensation (NO), full alignment sweep (FS), adaptive alignment sweep (AS), candidate elimination (CE), and adaptive alignment sweep combined with candidate elimination (AS+CE). For each (variant, configuration) pair, we run $100$ trials and increase the commit margin in $8$-byte steps until all trials recover the password successfully, allowing up to two retries per password to absorb transient measurement noise before recording a failure. \cref{tab:evaluation} reports the observed guess counts required for recovery, including those for the password length and terminator character. The NO and CE configurations presuppose a known winning alignment length, an assumption that the browser-based variant's noise floor does not support; we therefore mark the corresponding cells as \textit{n/a}.

\begin{table*}[tb]
    \centering
    \caption{Number of guesses required to recover an 8-character lowercase password across the three attack scenarios ($n = 100$). For each scenario, we evaluated no further compensation (NO), full alignment sweep (FS), adaptive alignment sweep (AS), candidate elimination (CE), and adaptive alignment sweep with candidate elimination (AS+CE). The high noise floor of browser-based injection mandates an alignment sweep; the corresponding NO and CE cells are therefore labeled \textit{n/a}. The commit margin was incremented in $8$-byte steps until all trials succeeded, allowing up to two retries per password to absorb transient measurement noise before recording a failure. Reported guess counts include those for the password length and termination characters.}
    \label{tab:evaluation}
    \small
    \setlength{\tabcolsep}{4pt}
    \begin{tabular}{lrrrrr|rrrrr|rrrrr}
        \toprule
         Variant & \multicolumn{5}{c}{Direct Plaintext Injection} & \multicolumn{5}{c}{Browser-Based Plaintext Injection} & \multicolumn{5}{c}{Ansible Password Recovery} \\
         ~ & \multicolumn{5}{c}{(\cref{sec:evaluation:sub:direct})} & \multicolumn{5}{c}{(\cref{sec:evaluation:sub:browser})} & \multicolumn{5}{c}{(\cref{sec:evaluation:sub:ansible})} \\
         & NO & FS & AS & CE & AS+CE & NO & FS & AS & CE & AS+CE & NO & FS & AS & CE & AS+CE \\ \midrule
         Margin & $8$ & $40$ & $16$ & $16$ & $80$ & \textit{n/a} & $48$ & $64$ & \textit{n/a} & $88$ & $8$ & $8$ & $8$ & $8$ & $8$ \\ \midrule
         Min. & $292$ & $14\,016$ & $2\,183$ & $763$ & $8\,717$ & \textit{n/a} & $27\,640$ & $36\,257$  & \textit{n/a} & $20\,514$ & $276$ & $2\,208$ & $696$ & $276$ & $696$  \\
         Max. & $1\,395$ & $34\,680$ & $8\,598$ & $2\,072$ & $21\,400$ & \textit{n/a} & $57\,400$ & $63\,776$  & \textit{n/a} & $33\,154$ & $474$ & $4\,368$ & $1\,983$ & $276$ & $696$ \\ \midrule
         Mean & $677.3$ & $20\,184.6$ & $4\,340.5$ & $1\,261.8$ & $13\,161.4$ & \textit{n/a} & $39\,660.6$ & $47\,366.0$  & \textit{n/a} & $27\,555.2$ & $284.0$ & $2\,467.7$ & $790.2$ & $276.0$ & $696.0$ \\
         Median & $621.5$ & $19\,488.0$ & $4\,048.0$ & $1\,260.0$ & $12\,995.0$ & \textit{n/a} & $39\,484.0$ & $47\,215.0$  & \textit{n/a} & $27\,619.5$ & $276.0$ & $2\,208.0$ & $696.0$ & $276.0$ & $696.0$ \\
         Std.~Dev. & $238.6$ & $3\,632.5$ & $1\,390.1$ & $239.2$ & $2\,182.8$ & \textit{n/a} & $6\,319.7$ & $5\,936.5$  & \textit{n/a} & $2\,757.2$ & $35.8$ & $528.3$ & $290.3$ & $0.0$ & $0.0$ \\
         \bottomrule
    \end{tabular}
\end{table*}

While these commit margins guarantee recovery, lower margins still succeed with substantial probability. At an $8$-byte margin, direct injection with a full alignment sweep (FS) recovers the password in $44$ of $100$ runs. Since lower commit margins also reduce the per-attack guess count, the values in \cref{tab:evaluation} are upper bounds for the case of guaranteed recovery rather than typical guess counts.

\subsection{SSH Implementations}
\label{sec:evaluation:sub:implementations}

To estimate how broadly our attack applies in practice, we compiled a selection of 33~SSH client implementations and applications based on the Quendi SSH comparison\footnote{\url{https://ssh-comparison.quendi.de/}} and tested each one manually for compression and port forwarding support. \cref{tab:implementations} reports the results. Four clients prefer compression by default; since SSH algorithm negotiation follows the client's preference order, these clients negotiate compression on every connection to a server that supports it. Many other clients offer compression as a non-preferred option, in which case the connection uses compression only if the server omits \texttt{none} from its algorithm list.

Almost all clients support local and remote port forwarding. Dynamic port forwarding appears mostly in GUI clients; libraries rarely offer it. Five clients bind local or remote forwards to all interfaces unless the user explicitly restricts them to localhost, which may expose the forward to direct injection (\cref{sec:evaluation:sub:direct}). However, no client both prefers compression and binds port forwards to all interfaces in its default configuration; default installations are therefore not directly vulnerable to a network attacker.

Overall, 29 clients are potentially affected by our generic attack, as they can be configured to be vulnerable. Only four clients either lack support for compression (Go x/crypto/ssh, Termius, and wolfSSH) or lack support for port forwarding (phpseclib) and thus cannot be configured to be vulnerable to our attack. However, we note that due to support for compression, phpseclib may still be vulnerable to other compression side-channel attacks.

\begin{table}
    \centering
    \caption{Selection of SSH client implementations and whether they are potentially affected by our attack. Our selection covers Windows~(\textnormal{\faWindows}), Linux~(\textnormal{\faLinux}), Android~(\textnormal{\faAndroid}), and clients available as libraries~(\textnormal{\faBook}). Clients configurable to be vulnerable are marked~(\textnormal{\affected}) in the Aff.\ column. For compression, we distinguish unsupported~(\textnormal{\unsupported}), disabled by default~(\textnormal{\partialsupport}), enabled by default~(\textnormal{\supported}), and preferred by default~(\textnormal{\supportedDefault}). For local~(L), remote~(R), and dynamic~(D) port forwarding, we distinguish unsupported~(\textnormal{\fwdunsupp}), supported~(\textnormal{\fwdsupp}), and supported with a public bind by default~(\textnormal{\fwdpublic}). @openssh.com is abbreviated to @.}
    \label{tab:implementations}
    \footnotesize
    \setlength{\tabcolsep}{4pt}
    \begin{tabular}{lcccccccc}
    \toprule
    Implementation & Version & OS & Aff. & \multicolumn{2}{c}{Compression} & \multicolumn{3}{c}{Port Fwd.} \\
    ~ & ~ & ~ & ~ & \texttt{zlib} & \texttt{zlib@} & L & R & D \\
    \midrule
    AbsoluteTelnet & 13.15 & \faWindows & \affected & \supported & \supported & \fwdpublic & \fwdpublic & \fwdpublic \\
    Ansible & 2.20.5 & \faLinux & \affected & \supportedDefault & \supportedDefault & \fwdsupp & \fwdsupp & \fwdsupp \\
    Apache MINA SSHD & 2.17.1 & \faBook & \affected & \partialsupport & \partialsupport & \fwdsupp & \fwdsupp & \fwdsupp \\
    AsyncSSH & 2.22.0 & \faBook & \affected & \partialsupport & \supported & \fwdsupp & \fwdsupp & \fwdsupp \\
    BitviseSSH & 9.59 & \faWindows & \affected & \supported & \unsupported & \fwdsupp & \fwdpublic & \fwdsupp \\
    ConnectBot & 1.10.6 & \faAndroid & \affected & \partialsupport & \partialsupport & \fwdsupp & \fwdpublic & \fwdsupp \\
    Dropbear & 2025.89 & \faLinux & \affected & \supportedDefault & \supportedDefault & \fwdsupp & \fwdsupp & \fwdunsupp \\
    Erlang OTP ssh & 5.5.2 & \faBook & \affected & \partialsupport & \supported & \fwdsupp & \fwdsupp & \fwdunsupp \\
    Go x/crypto/ssh & 0.50.0 & \faBook & \unaffected & \unsupported & \unsupported & \fwdsupp & \fwdsupp & \fwdunsupp \\
    JSch & 2.28.0 & \faBook & \affected & \partialsupport & \partialsupport & \fwdsupp & \fwdsupp & \fwdunsupp \\
    libssh & 0.12.0 & \faBook & \affected & \partialsupport & \supported & \fwdsupp & \fwdsupp & \fwdunsupp \\
    libssh2 & 1.11.1 & \faBook & \affected & \partialsupport & \partialsupport & \fwdsupp & \fwdsupp & \fwdunsupp \\
    MobaXTerm & 26.1.5456 & \faWindows & \affected & \supportedDefault & \supportedDefault & \fwdsupp & \fwdsupp & \fwdsupp \\
    Net:SSH & 7.3.2 & \faBook & \affected & \supported & \supported &  \fwdsupp & \fwdsupp & \fwdunsupp \\
    OpenSSH & 10.3p1 & \faLinux & \affected & \unsupported & \supported & \fwdsupp & \fwdsupp & \fwdsupp \\
    Paramiko & 4.0.0 & \faBook & \affected & \partialsupport & \partialsupport & \fwdsupp & \fwdsupp & \fwdunsupp \\
    PKIX-SSH & 18.0.2 & \faLinux & \affected & \supported & \supported & \fwdsupp & \fwdsupp & \fwdsupp \\
    phpseclib & 3.0.52 & \faBook & \unaffected & \supported & \supported & \fwdunsupp & \fwdunsupp & \fwdunsupp \\
    PrivX Desktop & 7.0.0.4175 & \faWindows & \affected & \supported & \unsupported & \fwdsupp & \fwdsupp & \fwdunsupp \\
    PuTTY & 0.83 & \faWindows & \affected & \supported & \supported & \fwdsupp & \fwdsupp & \fwdsupp \\
    russh & 0.60.1 & \faBook & \affected & \supported & \supported & \fwdsupp & \fwdsupp & \fwdunsupp \\
    SecureCRT & 9.7.1.3815 & \faWindows & \affected & \partialsupport & \partialsupport & \fwdsupp & \fwdsupp & \fwdsupp \\
    SSH.NET & 2025.1.0 & \faBook & \affected & \unsupported & \supported & \fwdsupp & \fwdsupp & \fwdsupp \\
    SSH2 & 1.17.0 & \faBook & \affected & \supported & \supported & \fwdsupp & \fwdsupp & \fwdunsupp \\
    SSHJ & 0.40.0 & \faBook & \affected & \partialsupport & \partialsupport & \fwdsupp & \fwdsupp & \fwdunsupp \\
    TeraTerm & 5.6.0 & \faWindows & \affected & \partialsupport & \partialsupport & \fwdsupp & \fwdsupp & \fwdsupp \\
    Termius & 9.37.5 & \faWindows~\faAndroid & \unaffected & \unsupported & \unsupported & \fwdsupp & \fwdsupp & \fwdsupp \\
    Termux (Dropbear) & 2026.02.11 & \faAndroid & \affected & \supportedDefault & \supportedDefault & \fwdsupp & \fwdsupp & \fwdunsupp \\
    Termux (OpenSSH) & 2026.02.11 & \faAndroid & \affected & \unsupported & \supported & \fwdsupp & \fwdsupp & \fwdsupp \\
    Twisted Conch & 25.5.0 & \faBook & \affected & \partialsupport & \unsupported & \fwdpublic & \fwdpublic & \fwdunsupp \\
    Win32-OpenSSH & 10.0p2 & \faWindows & \affected & \unsupported & \supported & \fwdsupp & \fwdsupp & \fwdsupp \\
    wolfSSH & 1.5.0 & \faBook & \unaffected & \unsupported & \unsupported & \fwdsupp & \fwdunsupp & \fwdunsupp \\
    Xshell & 8.0.0095 & \faWindows & \affected & \partialsupport & \partialsupport & \fwdpublic & \fwdsupp & \fwdpublic \\
    \bottomrule
    \end{tabular}
\end{table}

We also verified that support for compression is widespread among SSH servers. Per Censys\footnote{\url{https://censys.io/}} as of July 2026, 28.1M of 30.2M SSH servers (93\%) offer \texttt{zlib} or \texttt{zlib\allowbreak @openssh.com} in \sshKexInit, thus indicating support for compression. We conclude that use of compression is only lightly constrained by server support.

\section{Server-Side Attacks on X11 Forwarding}
\label{sec:x11}

Our evaluation above is based on a client-side attacker. In this section, we analyze how the same attack applies to a server-side attack scenario; we extend the threat model to distinguish the SSH server from the other, less privileged users on the same system. Unlike the three scenarios of \cref{sec:evaluation}, we did not fully evaluate this variant but only recovered the first byte of the MIT magic cookie to prove feasibility. First, we describe the necessary background on X11 and SSH forwarding of X11 connections. Then we describe how the attack preconditions can be fulfilled and what an attacker can gain from this configuration.

\subsection{X11 Authentication and Forwarding}
\label{sec:x11bg}

X11 is a server-client architecture for graphical desktop environments. The X11 server runs locally on the user's machine, utilizing the local graphics card and monitor as a high-resolution display. X11 clients are graphical applications that connect to the X11 server to open windows, update the window contents, receive keyboard and mouse events, interact with the clipboard, and more. While the server and client often run on the same machine, the architecture itself is networked, allowing clients to run on remote machines while the user interacts with them locally on the server.

Access to an X11 server is possible either through a UNIX domain socket or a network port, depending on the configuration. In either case, the client has to authenticate to the X11 server. The most common method is an \emph{MIT magic cookie}, a 16-byte secret stored in a file in the user's home directory and protected by OS access controls. To authenticate, the client reads out the secret and replays it at the beginning of the connection with a fixed 32-byte header:
\begin{figure}[H]
    \centering
    \includegraphics[width=\columnwidth]{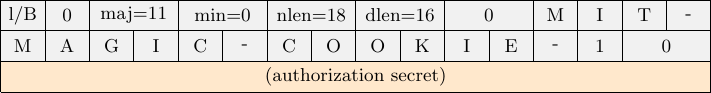}
    \Description{A block diagram showing a three-row data structure. The first row contains a sequence of fields labeled ``l/B'', ``0'', ``maj=11'', ``min=0'', ``nlen=18'', ``dlen=16'', ``0'', ``M'', ``I'', ``T'', and ``-''. The second row displays individual character cells that spell out the string ``M A G I C - C O O K I E - 1'', terminating with a cell containing ``0''. The third row consists of a single wide block spanning the entire width of the diagram, labeled with the text ``(authorization secret)''.}
\end{figure}

\paragraph{X11 Forwarding over SSH}

A user can give remote applications on an SSH server access to a local X11 server running on the user's system by using X11 forwarding. The SSH client requests the server to listen for incoming X11 client connections and open an \chName{x11} channel for each connection. The SSH client then in turn connects to the local X11 server and forwards all channel data to it. To allow remote X11 clients to authenticate, the SSH client includes the magic cookie in the original request to the server, which the server installs in the user's home directory on the remote server.

\paragraph{SSH Fake Cookies and Validation in OpenSSH}

While SSH implementations may forward the original magic cookie to the server, this is discouraged by the standard~\cite[Sect. 6.3.1]{rfc4254}, which instead encourages implementations to generate a fake cookie, which is installed on the remote machine. This fake cookie must then be validated by SSH and substituted with the original cookie on the client side. This transparent interposition protects the original cookie from exposure to remote machines. In OpenSSH, validation and substitution happen on the SSH client side. Thus, unauthenticated data is forwarded over the BPP before being blocked locally.

\subsection{Security Analysis of SSH X11 Forwarding}

SSH X11 forwarding can satisfy several preconditions of our adaptive compression attack without any additional port forwarding or shell session configured by the user. First, X11 forwarding creates secret-bearing channels. Each forwarded X11 connection begins with an authorization record that contains either the original X11 cookie or, in implementations such as OpenSSH, a fake cookie installed by the SSH server and validated by the SSH client. In both cases, the cookie appears at a fixed position after a known header in the first \sshChannelData message, which provides the known prefix required by the attack. Second, some X11 forwarding configurations also provide a plaintext-injection vector. If the forwarded X11 listener is exposed through a TCP port, an attacker who can connect to that port can send unauthenticated X11 connection data. In OpenSSH, X11 cookie validation happens at the SSH client. Consequently, the attacker's first \sshChannelData message is already compressed and transmitted over the connection before it is rejected. This gives the attacker a length-limited chosen-plaintext oracle on the same connection. Together, these properties make X11 forwarding a special case: the same SSH feature can provide both the secret-bearing channel and the injection channel. The attack still requires two additional conditions. The attacker must be able to observe the volume of encrypted traffic, and enough honest X11 client connections must occur while the cookie remains valid. Without a way to trigger victim X11 clients, the attacker has to wait for such connections.

\subsection{Attack Scenario}

Consider a victim who connects to a multi-user SSH server with
\texttt{ssh~-C~-X}, enabling both compression and X11 forwarding. No other SSH forwarding configuration is required. The attacker is another user on the same server. This attacker can connect to the server-side X11 forwarding listener and inject unauthenticated X11 connection attempts but does not have an on-path network vantage point between the victim and the SSH server.

This is a variant of the attacker model of \cref{sec:attack:sub:model}: instead of observing packets on the network path, the attacker obtains the same capability---the total bytes transmitted in a time interval---from host-level I/O statistics. For example, on Linux systems, an unprivileged user may be able to read transmitted-byte counters from the network interface via \texttt{/sys/class/net/eth0/statistics/tx\_bytes}. The attacker then applies the attack from \cref{sec:attack} to recover the fake X11 cookie installed for the victim's forwarded X11 session.

\paragraph{Discussion}

Recovering the cookie allows the attacker to connect attacker-controlled X11 clients through the victim's forwarded X11 channel. Depending on X11 security settings and extensions, this may allow the attacker to open windows, display content, steal focus, or even interact with the clipboard or observe input or window contents. The attacker might also be able to attack the X11 server implementation directly, e.g., by exploiting memory safety bugs. With the original cookie, a local network attacker on the victim's system could also connect to the X11 server directly.

This scenario differs from the attacks in \cref{sec:evaluation} in three ways. First, the targeted secret is defined by the SSH Connection Protocol itself: X11 forwarding requires an authorization cookie at the start of each forwarded X11 connection. Second, the relevant traffic direction is reversed: the server opens the \chName{x11} channel, and the secret flows from the server side toward the client. As noted above, client-side validation admits unauthenticated data into the compression context.

\section{Related Work}
\label{sec:related}

\paragraph{SSH Protocol Security}

Previous work has largely focused on the security of the SSH key exchange~\cite{CCS:RHSH23,CCS:BMBS25,CCS:BDKSS14,CSF:BlaJac24,SP:BDMX25,NDSS:BhaLeu16}, user authentication~\cite{CCS:BBRSS25}, the transport layer~\cite{CCS:ADHP16,CCS:BelKohNam02,EC:PatWat10,SP:AlbPatWat09,USENIX:BauBriSch24}, and keystroke timing leakage in interactive sessions~\cite{USENIX:SonWagTia01}.

\paragraph{Compression Side Channels}

Kelsey first observed that compressing plaintext before length-preserving encryption can leak information through ciphertext lengths~\cite{FSE:Kelsey02}. Subsequent work demonstrated practical attacks in several settings, including TLS compression in CRIME~\cite{CRIME}, SPDY4~\cite{TanNahata2013PETAL}, HTTP response compression in BREACH~\cite{prado2013breach}, XMPP~\cite{alkemade2014xmpp}, encrypted VoIP~\cite{SP:WMSM11}, video streams~\cite{USENIX:SchShmTro17}, OpenVPN~\cite{nafeez2018voracle}, databases~\cite{SP:HogMicEsk23,ACNS:BouMicEsk25}, and cloud storage~\cite{SP:FPNNAR24,USENIX:FNANR24}. The HEIST~\cite{vanhoef2016heist} and TIME~\cite{beery2013perfectcrime} attacks improved the CRIME attack by replacing the MitM eavesdropper with a timing-based side channel in the browser. The possibility of similar attacks in SSH was previously discussed on the OpenSSH mailing list in 2014, but without considering channel multiplexing.\footnote{\url{https://lists.mindrot.org/pipermail/openssh-unix-dev/2014-November/033176.html}} These attacks have in common that attacker-controlled input can change the compressed length of nearby secret-bearing data or vice versa.

\paragraph{Security of Channel Multiplexing}

HTTP/2 and HTTP/3 also use multiplexing but restrict compression to headers. Their header-compression schemes, HPACK~\cite{rfc7541} and QPACK~\cite{rfc9204}, explicitly support excluding secret header values from compression to mitigate adaptive compression side channels. Aside from compression, Van Goethem et al.~\cite{USENIX:GPJV20} showed how to exploit HTTP/2 request and response multiplexing in concurrency-based timing attacks.

\section{Discussion}
\label{sec:discussion}

The root cause of our attack is structural: SSH maintains one compression context per connection direction, and within each direction, every byte from every multiplexed channel passes through the same zlib search buffer before encryption. We have shown this in \cref{sec:analysis} and exploited it in \cref{sec:evaluation}; what is worth restating is where the attack surface originates from. SSH splits its protocol across multiple specifications: RFC 4253~\cite{rfc4253} defines the BPP, including compression, while RFC 4254~\cite{rfc4254} defines the Connection Protocol, including channel multiplexing. Each describes its layer in isolation: the Connection Protocol presents channels as independent streams, while the BPP, sitting beneath it, treats the bytes flowing through as opaque and is unaware that channels exist. Neither specification states that channels share a compression context, because the question does not arise within either layer alone. The attack targets exactly this gap. It is also why prior work did not find it: CRIME~\cite{CRIME} and BREACH~\cite{prado2013breach} operate within a single HTTPS connection, where the shared compression context is obvious; SSH's channel multiplexing makes that sharing easy to miss. The community response also differed: TLS removed compression entirely after CRIME; SSH did not. SSH compression remains enabled by default in several clients (\cref{tab:implementations}) and is recommended in vendor documentation for slow links. There has been no CRIME-class incident in the SSH ecosystem to force the same hard removal; our work raises the question of whether one is required or whether the design choice can be revisited proactively.

\subsection{Feasibility and Limitations}
\label{sec:discussion:sub:feasability}

Our results are subject to three kinds of limitations: what our analysis establishes about deployed systems, the capabilities we grant the attacker in our evaluation, and the evaluated cipher configuration.

\paragraph{Exposure of Deployed Systems} 

We establish the feasibility of an adaptive compression attack under specific configurations, but we do not measure their prevalence. The required combination---compression negotiated, multiple channels multiplexed in the same direction, one secret-bearing and one offering a partially chosen-plaintext oracle, and repeated secret transmission---is configuration-dependent, and we have no insight into its joint occurrence in real deployments; quantifying exposure remains an open question. Where the combination does occur, the quality of the resulting oracle still varies: noise, padding, application behavior, and, in the browser-based scenario, the restrictions discussed in \cref{sec:evaluation:sub:browser} can each degrade the oracle or remove it altogether (cf. \cref{sec:discussion:sub:noise}).

\paragraph{Active Triggering}

For our evaluation in \cref{sec:evaluation}, we grant the attacker the ability to actively trigger the transmission of the secret. This is realistic in some scenarios---for example, for an unattended client that automatically reconnects and reauthenticates after transient network failures. A network attacker can induce such failures, e.g., using TCP RST packets, causing repeated authentication attempts without user interaction. An attacker without this ability is restricted in two ways: the number of secret transmissions can no longer be increased at will and is limited to those that occur naturally, and the primary mechanism for synchronizing the attack with the transmission of the secret is lost. The first restriction affects the runtime of the attack rather than the number of guesses it requires. In \cref{tab:evaluation}, we therefore compare runs by the number of guesses rather than by runtime. Under ideal conditions, each guess requires one secret transmission, so the guess count is also the number of transmissions required, and the average runtime is the product of the guess count and the average time per guess. An attacker who has to wait for transmissions that occur naturally pays a higher average time per guess, and the runtime grows in proportion. For the direct plaintext injection of \cref{sec:evaluation:sub:direct}, with a median of $621.5$ guesses in the NO configuration, an average of one minute per guess yields a runtime of about ten hours, whereas one hour per guess extends it to roughly 26 days. The second restriction affects the number of guesses. Without synchronization, guesses made while the secret is outside the LZ77 compression context do not contribute to leaking it, which raises the number of guesses needed to recover the secret. Synchronization can instead be achieved, for example, when the secret is sent at connection start.

\paragraph{Cipher Dependence}

Our results are tuned to ChaCha20-Poly1305, specifically through the alignment sweep, which follows its padding granularity. AES-based modes pad to 16~bytes; the attack still applies, but the sweep grows with the block length of the cipher, and with it, the number of guesses required increases linearly.

\subsection{Impact of Noise}
\label{sec:discussion:sub:noise}

Our attack relies on measuring the difference in compression length between correct and wrong guesses. Noise hinders this measurement, requiring several measurements per candidate byte to average the noise out. In high-noise scenarios, detecting the difference may become impossible altogether. We employ noise compensation strategies described in \cref{sec:attack:sub:optimization}, which allow for successful recovery even in the presence of ample noise.

There are different sources of noise but with a similar effect. The SSH protocol uses a shared compression context across all of its channels. As a result, noise on a different, possibly unrelated channel has the same effect as noise within the guess or secret channel. Our evaluation already covers a scenario with significant noise: the browser-based plaintext injection described in \cref{sec:evaluation:sub:browser}. There, the noise comes from additional HTTP headers and from the random data prepended to force a static Huffman block. This noise resides in the guess channel and, by the argument above, is equivalent to similar noise on a different channel---for example, a parallel HTTP request in the direct injection scenario of \cref{sec:evaluation:sub:direct}.

There is a hard limit on how much noise still allows the secret to be recovered: if throughput is so high that the attacker cannot reliably inject its guess close to the secret, the attack cannot succeed.

\subsection{Countermeasures}
\label{sec:countermeasures}

The most effective countermeasure is also the simplest: disable SSH compression. Without compression, channel multiplexing and port forwarding pose no compression side-channel risk. If compression must remain enabled---for example, on bandwidth-constrained links---users can isolate sensitive sessions on their SSH connection: instead of using multiplex port forwarding with channels that may carry secrets, they can start a fresh connection for any sensitive operation. Note that delaying compression until after user authentication (e.g., \texttt{zlib@openssh.com}) does not mitigate our attack because the Connection Protocol runs after user authentication.

SSH implementations could encourage strong isolation by default. They could warn when compression and port forwarding are enabled together, document the cross-channel scope of the compression context, and default to opening fresh SSH connections rather than sharing existing ones where feasible. OpenSSH's \texttt{ControlMaster} feature is the most prominent case to address: it transparently multiplexes new sessions over an existing connection, extending the shared compression context to traffic the user may not have anticipated. We recommend leaving \texttt{ControlMaster} off by default and documenting its interaction with compression. For the browser variant, PNA reduces the attack surface by gating public-to-private cross-origin requests. However, as discussed in \cref{sec:evaluation:sub:browser}, the preflight still carries attacker-controllable bytes in the URL path and headers, reducing but not eliminating the oracle.

A more in-depth countermeasure that would still allow the use of compression within the SSH protocol would be to dedicate separate compression states to each channel. However, this approach is not straightforward, as SSH, according to RFC~4253~\cite[Sect. 6.2]{rfc4253}, compresses the entire payload field of the binary packet, which includes the channel ID. Therefore, a peer must decompress the payload first before being able to distinguish messages sent to different channels. It is therefore easy to see that any compression with a per-channel compression context cannot be implemented in accordance with the original specification. However, previous algorithms such as \texttt{zlib@openssh.com} show that there is a precedent for compression algorithms that deviate from RFC~4253's behavior while being negotiated through the vendor-extension namespace the specification provides. Thus, deviations from the original RFCs could be acceptable when they are appropriately scoped.

We therefore propose a solution that we call \emph{selective, per-channel compression}: depending on the negotiation of this new compression algorithm, only channel messages sent after the channel-open handshake (message IDs 93 to 100) have their message data compressed, excluding the message ID and channel ID fields. For all other messages, compression is not used. Whenever a new channel is opened, both peers allocate a new compression context and associate the compression context with their respective channel IDs. When receiving a message, the peer processes the (uncompressed) message ID, determining whether the message is compressed or not. If it is, the peer then continues by choosing the correct compression context and decompressing the data using the (uncompressed) channel ID. Whenever the channel is closed, the compression state for that channel is discarded. To avoid dependencies between encryption states, all compression states must be reinitialized following a key re-exchange. An interesting side effect of the described selective, per-channel compression is delayed compression; that is, no compression is used before successful user authentication. The downside of this proposal is a higher memory cost for compression contexts, while bandwidth costs may improve through better locality of similar data or worsen for short-lived channels.

At the protocol level, we will engage with the IETF sshm working group to raise awareness and discuss mitigations. The most direct path follows TLS's response to CRIME: deprecate compression in the SSH standard. Whether the community takes that step, relies on warnings and more secure defaults, or even supports new compression algorithms, such as selective, per-channel compression, is a question we believe should be discussed proactively.

\section{Conclusion}

In our work, we have analyzed the security of the SSH Connection Protocol regarding cross-channel state dependencies at the SSH Binary Packet Protocol layer. Answering RQ1, channel multiplexing creates a security-relevant interaction between channels: the shared compression context couples the compressed lengths of secret-bearing and attacker-controlled packets, allowing an eavesdropper with a partially chosen-plaintext oracle to recover those secrets.

Answering RQ2, we have examined which SSH channel types are secret-bearing and which can serve as plaintext injection vectors. As special cases, X11 forwarding is always secret-bearing because it carries an authentication token; SSH agent forwarding is not secret-bearing by design. The remaining channel types are application-defined and therefore potentially secret-bearing, with TCP port forwarding providing a high-bandwidth, low-noise injection channel for a network attacker. Most importantly, all SSH channels share a single compression context at the BPP layer; this is the cross-channel dependency our attack exploits.

Answering RQ3, we have identified adaptive compression attacks via TCP port forwarding as a major composition risk in the SSH protocol and presented a general attack strategy that recovers a secret under certain conditions. Critically, this attack differs from the CRIME and BREACH attacks on TLS, which require the secret-bearing and injection channels to be the same; in SSH, the two can be different channels as long as they share the same connection.

We have verified our findings with a proof-of-concept evaluation in three different real-world configurations and demonstrated the attack's sensitivity to noise and alignment recovery. Our attack is practical, recovering an 8-character secret over a 26-letter alphabet in $276$ (Ansible), $1\,260$ (direct injection), and $27\,619.5$ (browser-based) guesses (median over 100 trials with candidate elimination, plus an adaptive alignment sweep for the browser variant; \cref{tab:evaluation}). The latter requires more guesses because of additional noise in the combined eavesdropper-and-web-attacker model we bring to SSH.

Our work identifies SSH channel multiplexing as a security-relevant boundary, demonstrates a concrete attack that exploits it, and guides users, vendors, and standardization committees toward safer configurations, implementations, and protocol standards.

\section*{Generative AI Usage}

We used ChatGPT, Gemini, and Claude during the preparation of this work. In particular, we used the responses from these models to deeply reflect on the structure of our text and revise individual paragraphs, aiming for improved readability and coherence while ensuring an accurate representation of our research. In addition, we continuously used LanguageTool for editorial changes.

The proof-of-concept implementations for the three attack scenarios in \cref{sec:evaluation} were implemented with Claude Code. We provided the generic attack description, attacker model, scenario descriptions, and an explicit requirement to avoid implementation shortcuts. We then manually reviewed the code to ensure the attacker model and scenarios were implemented accurately without unexpected shortcuts. Dedicated integration tests verified correctness by recovering passwords end-to-end. Together with our code review, these tests ensure the figures in \cref{tab:evaluation} are accurate.

\bibliographystyle{ACM-Reference-Format}
\balance
\bibliography{bib/abbrev2.bib, bib/crypto.bib, bib/rfc.bib, bib/paper.bib}


\begin{thebibliography}{56}


\ifx \showCODEN    \undefined \def \showCODEN     #1{\unskip}     \fi
\ifx \showISBNx    \undefined \def \showISBNx     #1{\unskip}     \fi
\ifx \showISBNxiii \undefined \def \showISBNxiii  #1{\unskip}     \fi
\ifx \showISSN     \undefined \def \showISSN      #1{\unskip}     \fi
\ifx \showLCCN     \undefined \def \showLCCN      #1{\unskip}     \fi
\ifx \shownote     \undefined \def \shownote      #1{#1}          \fi
\ifx \showarticletitle \undefined \def \showarticletitle #1{#1}   \fi
\ifx \showURL      \undefined \def \showURL       {\relax}        \fi
\providecommand\bibfield[2]{#2}
\providecommand\bibinfo[2]{#2}
\providecommand\natexlab[1]{#1}
\providecommand\showeprint[2][]{arXiv:#2}

\bibitem[Albrecht et~al\mbox{.}(2016)]%
        {CCS:ADHP16}
\bibfield{author}{\bibinfo{person}{Martin~R. Albrecht}, \bibinfo{person}{Jean~Paul Degabriele}, \bibinfo{person}{Torben~Brandt Hansen}, {and} \bibinfo{person}{Kenneth~G. Paterson}.} \bibinfo{year}{2016}\natexlab{}.
\newblock \showarticletitle{A Surfeit of {SSH} Cipher Suites}. In \bibinfo{booktitle}{\emph{ACM CCS 2016}}, \bibfield{editor}{\bibinfo{person}{Edgar~R. Weippl}, \bibinfo{person}{Stefan Katzenbeisser}, \bibinfo{person}{Christopher Kruegel}, \bibinfo{person}{Andrew~C. Myers}, {and} \bibinfo{person}{Shai Halevi}} (Eds.). \bibinfo{publisher}{{ACM} Press}, \bibinfo{address}{Vienna, Austria}, \bibinfo{pages}{1480--1491}.
\newblock
\href{https://doi.org/10.1145/2976749.2978364}{doi:\nolinkurl{10.1145/2976749.2978364}}


\bibitem[Albrecht et~al\mbox{.}(2009)]%
        {SP:AlbPatWat09}
\bibfield{author}{\bibinfo{person}{Martin~R. Albrecht}, \bibinfo{person}{Kenneth~G. Paterson}, {and} \bibinfo{person}{Gaven~J. Watson}.} \bibinfo{year}{2009}\natexlab{}.
\newblock \showarticletitle{Plaintext Recovery Attacks against {SSH}}. In \bibinfo{booktitle}{\emph{2009 {IEEE} Symposium on Security and Privacy}}. \bibinfo{publisher}{{IEEE} Computer Society Press}, \bibinfo{address}{Oakland, CA, USA}, \bibinfo{pages}{16--26}.
\newblock
\href{https://doi.org/10.1109/SP.2009.5}{doi:\nolinkurl{10.1109/SP.2009.5}}


\bibitem[Alkemade(2014)]%
        {alkemade2014xmpp}
\bibfield{author}{\bibinfo{person}{Thijs Alkemade}.} \bibinfo{year}{2014}\natexlab{}.
\newblock \bibinfo{title}{{HTTPS} Attacks and {XMPP} 2: {CRIME} \& {BREACH}}.
\newblock
\urldef\tempurl%
\url{https://blog.thijsalkema.de/blog/2014/08/07/https-attacks-and-xmpp-2-crime-and-breach/}
\showURL{%
\tempurl}
\newblock
\shownote{Accessed: 2026-04-24}.


\bibitem[{Amazon Web Services}(2026)]%
        {aws-rds-ssh-tunnel}
\bibfield{author}{\bibinfo{person}{{Amazon Web Services}}.} \bibinfo{year}{2026}\natexlab{}.
\newblock \bibinfo{title}{{Amazon} Prescriptive Guidance---Connect to a {PostgreSQL} {DB} instance in {Amazon} {RDS} by using an {SSH} tunnel in {pgAdmin}}.
\newblock
\urldef\tempurl%
\url{https://docs.aws.amazon.com/prescriptive-guidance/latest/patterns/connect-by-using-an-ssh-tunnel-in-pgadmin.html}
\showURL{%
\tempurl}
\newblock
\shownote{Accessed: 2026-04-24}.


\bibitem[B{\"a}umer et~al\mbox{.}(2025a)]%
        {CCS:BBRSS25}
\bibfield{author}{\bibinfo{person}{Fabian B{\"a}umer}, \bibinfo{person}{Marcus Brinkmann}, \bibinfo{person}{Maximilian Radoy}, \bibinfo{person}{J{\"o}rg Schwenk}, {and} \bibinfo{person}{Juraj Somorovsky}.} \bibinfo{year}{2025}\natexlab{a}.
\newblock \showarticletitle{On the Security of {SSH} Client Signatures}. In \bibinfo{booktitle}{\emph{ACM CCS 2025}}. \bibinfo{publisher}{{ACM} Press}, \bibinfo{pages}{4619--4633}.
\newblock
\href{https://doi.org/10.1145/3719027.3765079}{doi:\nolinkurl{10.1145/3719027.3765079}}


\bibitem[B{\"a}umer et~al\mbox{.}(2024)]%
        {USENIX:BauBriSch24}
\bibfield{author}{\bibinfo{person}{Fabian B{\"a}umer}, \bibinfo{person}{Marcus Brinkmann}, {and} \bibinfo{person}{J{\"o}rg Schwenk}.} \bibinfo{year}{2024}\natexlab{}.
\newblock \showarticletitle{Terrapin Attack: Breaking {SSH} Channel Integrity By Sequence Number Manipulation}. In \bibinfo{booktitle}{\emph{USENIX Security 2024}}, \bibfield{editor}{\bibinfo{person}{Davide Balzarotti} {and} \bibinfo{person}{Wenyuan Xu}} (Eds.). \bibinfo{publisher}{{USENIX} Association}, \bibinfo{address}{Philadelphia, PA, USA}.
\newblock
\urldef\tempurl%
\url{https://www.usenix.org/conference/usenixsecurity24/presentation/b%C3%A4umer}
\showURL{%
\tempurl}


\bibitem[B{\"a}umer et~al\mbox{.}(2025b)]%
        {CCS:BMBS25}
\bibfield{author}{\bibinfo{person}{Fabian B{\"a}umer}, \bibinfo{person}{Marcel Maehren}, \bibinfo{person}{Marcus Brinkmann}, {and} \bibinfo{person}{J{\"o}rg Schwenk}.} \bibinfo{year}{2025}\natexlab{b}.
\newblock \showarticletitle{Finding {SSH} Strict Key Exchange Violations by State Learning}. In \bibinfo{booktitle}{\emph{ACM CCS 2025}}. \bibinfo{publisher}{{ACM} Press}, \bibinfo{pages}{246--260}.
\newblock
\href{https://doi.org/10.1145/3719027.3765208}{doi:\nolinkurl{10.1145/3719027.3765208}}


\bibitem[Be'ery and Shulman(2013)]%
        {beery2013perfectcrime}
\bibfield{author}{\bibinfo{person}{Tal Be'ery} {and} \bibinfo{person}{Amichai Shulman}.} \bibinfo{year}{2013}\natexlab{}.
\newblock \bibinfo{title}{A Perfect {CRIME}? Only {TIME} Will Tell}.
\newblock \bibinfo{howpublished}{Black Hat Europe 2013}.
\newblock
\urldef\tempurl%
\url{https://media.blackhat.com/eu-13/briefings/Beery/bh-eu-13-a-perfect-crime-beery-wp.pdf}
\showURL{%
\tempurl}
\newblock
\shownote{White paper, Accessed: 2026-04-24}.


\bibitem[Bellare et~al\mbox{.}(2002)]%
        {CCS:BelKohNam02}
\bibfield{author}{\bibinfo{person}{Mihir Bellare}, \bibinfo{person}{Tadayoshi Kohno}, {and} \bibinfo{person}{Chanathip Namprempre}.} \bibinfo{year}{2002}\natexlab{}.
\newblock \showarticletitle{Authenticated Encryption in {SSH}: Provably Fixing The {SSH} Binary Packet Protocol}. In \bibinfo{booktitle}{\emph{ACM CCS 2002}}, \bibfield{editor}{\bibinfo{person}{Vijayalakshmi Atluri}} (Ed.). \bibinfo{publisher}{{ACM} Press}, \bibinfo{address}{Washington, DC, USA}, \bibinfo{pages}{1--11}.
\newblock
\href{https://doi.org/10.1145/586110.586112}{doi:\nolinkurl{10.1145/586110.586112}}


\bibitem[Ben{\v c}ina et~al\mbox{.}(2025)]%
        {SP:BDMX25}
\bibfield{author}{\bibinfo{person}{Benjamin Ben{\v c}ina}, \bibinfo{person}{Benjamin Dowling}, \bibinfo{person}{Varun Maram}, {and} \bibinfo{person}{Keita Xagawa}.} \bibinfo{year}{2025}\natexlab{}.
\newblock \showarticletitle{Post-Quantum Cryptographic Analysis of {SSH}}. In \bibinfo{booktitle}{\emph{2025 {IEEE} Symposium on Security and Privacy}}. \bibinfo{publisher}{{IEEE} Computer Society Press}, \bibinfo{address}{San Francisco, CA, USA}, \bibinfo{pages}{595--613}.
\newblock
\href{https://doi.org/10.1109/SP61157.2025.00126}{doi:\nolinkurl{10.1109/SP61157.2025.00126}}


\bibitem[Bergsma et~al\mbox{.}(2014)]%
        {CCS:BDKSS14}
\bibfield{author}{\bibinfo{person}{Florian Bergsma}, \bibinfo{person}{Benjamin Dowling}, \bibinfo{person}{Florian Kohlar}, \bibinfo{person}{J{\"o}rg Schwenk}, {and} \bibinfo{person}{Douglas Stebila}.} \bibinfo{year}{2014}\natexlab{}.
\newblock \showarticletitle{Multi-Ciphersuite Security of the {Secure} {Shell} ({SSH}) Protocol}. In \bibinfo{booktitle}{\emph{ACM CCS 2014}}, \bibfield{editor}{\bibinfo{person}{Gail-Joon Ahn}, \bibinfo{person}{Moti Yung}, {and} \bibinfo{person}{Ninghui Li}} (Eds.). \bibinfo{publisher}{{ACM} Press}, \bibinfo{address}{Scottsdale, AZ, USA}, \bibinfo{pages}{369--381}.
\newblock
\href{https://doi.org/10.1145/2660267.2660286}{doi:\nolinkurl{10.1145/2660267.2660286}}


\bibitem[Bhargavan and Leurent(2016)]%
        {NDSS:BhaLeu16}
\bibfield{author}{\bibinfo{person}{Karthikeyan Bhargavan} {and} \bibinfo{person}{Ga{\"e}tan Leurent}.} \bibinfo{year}{2016}\natexlab{}.
\newblock \showarticletitle{Transcript Collision Attacks: Breaking Authentication in {TLS}, {IKE} and {SSH}}. In \bibinfo{booktitle}{\emph{NDSS~2016}}. \bibinfo{publisher}{The Internet Society}, \bibinfo{address}{San Diego, CA, USA}.
\newblock
\href{https://doi.org/10.14722/ndss.2016.23418}{doi:\nolinkurl{10.14722/ndss.2016.23418}}


\bibitem[Bider(2018)]%
        {rfc8308}
\bibfield{author}{\bibinfo{person}{Denis Bider}.} \bibinfo{year}{2018}\natexlab{}.
\newblock \bibinfo{title}{{Extension Negotiation in the Secure Shell (SSH) Protocol}}.
\newblock \bibinfo{howpublished}{RFC 8308}.
\newblock
\href{https://doi.org/10.17487/RFC8308}{doi:\nolinkurl{10.17487/RFC8308}}


\bibitem[Blanchet and Jacomme(2024)]%
        {CSF:BlaJac24}
\bibfield{author}{\bibinfo{person}{Bruno Blanchet} {and} \bibinfo{person}{Charlie Jacomme}.} \bibinfo{year}{2024}\natexlab{}.
\newblock \showarticletitle{Post-Quantum Sound {CryptoVerif} and Verification of Hybrid {TLS} and {SSH} Key-Exchanges}. In \bibinfo{booktitle}{\emph{CSF 2024 Computer Security Foundations Symposium}}. \bibinfo{publisher}{{IEEE} Computer Society Press}, \bibinfo{address}{Enschede, The Netherlands}, \bibinfo{pages}{543--556}.
\newblock
\href{https://doi.org/10.1109/CSF61375.2024.00050}{doi:\nolinkurl{10.1109/CSF61375.2024.00050}}


\bibitem[Bourassa et~al\mbox{.}(2025)]%
        {ACNS:BouMicEsk25}
\bibfield{author}{\bibinfo{person}{Britney Bourassa}, \bibinfo{person}{Yan Michalevsky}, {and} \bibinfo{person}{Saba Eskandarian}.} \bibinfo{year}{2025}\natexlab{}.
\newblock \showarticletitle{G-{DBREACH} Attacks: Algorithmic Techniques for Faster and Stronger Compression Side Channels}. In \bibinfo{booktitle}{\emph{ACNS 2025, Part~II}} \emph{(\bibinfo{series}{{LNCS}})}. \bibinfo{publisher}{Springer, Cham, Switzerland}, \bibinfo{pages}{24--48}.
\newblock
\href{https://doi.org/10.1007/978-3-031-95764-2_2}{doi:\nolinkurl{10.1007/978-3-031-95764-2_2}}


\bibitem[Dai(2002)]%
        {weidai2002}
\bibfield{author}{\bibinfo{person}{Wei Dai}.} \bibinfo{year}{2002}\natexlab{}.
\newblock \bibinfo{title}{{email to IETF mailing list}}.
\newblock \bibinfo{howpublished}{\url{https://www.ietf.org/ietf-ftp/ietf-mail-archive/secsh/2002-02.mail}}.
\newblock
\urldef\tempurl%
\url{https://www.ietf.org/ietf-ftp/ietf-mail-archive/secsh/2002-02.mail}
\showURL{%
\tempurl}
\newblock
\shownote{Accessed: 2026-04-24}.


\bibitem[Deutsch(1996a)]%
        {rfc1951}
\bibfield{author}{\bibinfo{person}{L.~Peter Deutsch}.} \bibinfo{year}{1996}\natexlab{a}.
\newblock \bibinfo{title}{{DEFLATE Compressed Data Format Specification version 1.3}}.
\newblock \bibinfo{howpublished}{RFC 1951}.
\newblock
\href{https://doi.org/10.17487/RFC1951}{doi:\nolinkurl{10.17487/RFC1951}}


\bibitem[Deutsch(1996b)]%
        {rfc1952}
\bibfield{author}{\bibinfo{person}{L.~Peter Deutsch}.} \bibinfo{year}{1996}\natexlab{b}.
\newblock \bibinfo{title}{{GZIP file format specification version 4.3}}.
\newblock \bibinfo{howpublished}{RFC 1952}.
\newblock
\href{https://doi.org/10.17487/RFC1952}{doi:\nolinkurl{10.17487/RFC1952}}


\bibitem[Deutsch and loup Gailly(1996)]%
        {rfc1950}
\bibfield{author}{\bibinfo{person}{L.~Peter Deutsch} {and} \bibinfo{person}{Jean loup Gailly}.} \bibinfo{year}{1996}\natexlab{}.
\newblock \bibinfo{title}{{ZLIB Compressed Data Format Specification version 3.3}}.
\newblock \bibinfo{howpublished}{RFC 1950}.
\newblock
\href{https://doi.org/10.17487/RFC1950}{doi:\nolinkurl{10.17487/RFC1950}}


\bibitem[F{\'a}brega et~al\mbox{.}(2024a)]%
        {USENIX:FNANR24}
\bibfield{author}{\bibinfo{person}{Andr{\'e}s F{\'a}brega}, \bibinfo{person}{Armin Namavari}, \bibinfo{person}{Rachit Agarwal}, \bibinfo{person}{Ben Nassi}, {and} \bibinfo{person}{Thomas Ristenpart}.} \bibinfo{year}{2024}\natexlab{a}.
\newblock \showarticletitle{Exploiting Leakage in Password Managers via Injection Attacks}. In \bibinfo{booktitle}{\emph{USENIX Security 2024}}, \bibfield{editor}{\bibinfo{person}{Davide Balzarotti} {and} \bibinfo{person}{Wenyuan Xu}} (Eds.). \bibinfo{publisher}{{USENIX} Association}, \bibinfo{address}{Philadelphia, PA, USA}.
\newblock
\urldef\tempurl%
\url{https://www.usenix.org/conference/usenixsecurity24/presentation/fabrega}
\showURL{%
\tempurl}


\bibitem[F{\'a}brega et~al\mbox{.}(2024b)]%
        {SP:FPNNAR24}
\bibfield{author}{\bibinfo{person}{Andr{\'e}s F{\'a}brega}, \bibinfo{person}{Carolina~Ortega P{\'e}rez}, \bibinfo{person}{Armin Namavari}, \bibinfo{person}{Ben Nassi}, \bibinfo{person}{Rachit Agarwal}, {and} \bibinfo{person}{Thomas Ristenpart}.} \bibinfo{year}{2024}\natexlab{b}.
\newblock \showarticletitle{Injection Attacks Against End-to-End Encrypted Applications}. In \bibinfo{booktitle}{\emph{2024 {IEEE} Symposium on Security and Privacy}}. \bibinfo{publisher}{{IEEE} Computer Society Press}, \bibinfo{address}{San Francisco, CA, USA}, \bibinfo{pages}{2648--2665}.
\newblock
\href{https://doi.org/10.1109/SP54263.2024.00082}{doi:\nolinkurl{10.1109/SP54263.2024.00082}}


\bibitem[Gluck et~al\mbox{.}(2013)]%
        {prado2013breach}
\bibfield{author}{\bibinfo{person}{Yoel Gluck}, \bibinfo{person}{Neal Harris}, {and} \bibinfo{person}{Angelo Prado}.} \bibinfo{year}{2013}\natexlab{}.
\newblock \bibinfo{title}{{BREACH}: Reviving the {CRIME} Attack}.
\newblock \bibinfo{howpublished}{Black Hat USA 2013}.
\newblock
\urldef\tempurl%
\url{https://media.blackhat.com/us-13/US-13-Prado-SSL-Gone-in-30-seconds-A-BREACH-beyond-CRIME-WP.pdf}
\showURL{%
\tempurl}
\newblock
\shownote{White paper, Accessed: 2026-04-24}.


\bibitem[{Google Cloud}(2026a)]%
        {google-dms}
\bibfield{author}{\bibinfo{person}{{Google Cloud}}.} \bibinfo{year}{2026}\natexlab{a}.
\newblock \bibinfo{title}{{Google} Cloud Documentation---Configure connectivity using a reverse SSH tunnel}.
\newblock
\urldef\tempurl%
\url{https://docs.cloud.google.com/database-migration/docs/mysql/configure-connectivity-reverse-ssh-tunnel}
\showURL{%
\tempurl}
\newblock
\shownote{Accessed: 2026-04-24}.


\bibitem[{Google Cloud}(2026b)]%
        {google-ds}
\bibfield{author}{\bibinfo{person}{{Google Cloud}}.} \bibinfo{year}{2026}\natexlab{b}.
\newblock \bibinfo{title}{Google Cloud Documentation---Forward SSH tunnel}.
\newblock
\urldef\tempurl%
\url{https://docs.cloud.google.com/datastream/docs/ssh-tunnel}
\showURL{%
\tempurl}
\newblock
\shownote{Accessed: 2026-04-24}.


\bibitem[{Google Cloud}(2026c)]%
        {google-looker}
\bibfield{author}{\bibinfo{person}{{Google Cloud}}.} \bibinfo{year}{2026}\natexlab{c}.
\newblock \bibinfo{title}{{Google} Cloud Documentation---Using an SSH tunnel}.
\newblock
\urldef\tempurl%
\url{https://docs.cloud.google.com/looker/docs/using-an-ssh-tunnel}
\showURL{%
\tempurl}
\newblock
\shownote{Accessed: 2026-04-24}.


\bibitem[Harding(2019)]%
        {autossh}
\bibfield{author}{\bibinfo{person}{Carson Harding}.} \bibinfo{year}{2019}\natexlab{}.
\newblock \bibinfo{title}{autossh}.
\newblock
\urldef\tempurl%
\url{https://www.harding.motd.ca/autossh/}
\showURL{%
\tempurl}
\newblock
\shownote{Accessed: 2026-04-24}.


\bibitem[Hogan et~al\mbox{.}(2023)]%
        {SP:HogMicEsk23}
\bibfield{author}{\bibinfo{person}{Mathew Hogan}, \bibinfo{person}{Yan Michalevsky}, {and} \bibinfo{person}{Saba Eskandarian}.} \bibinfo{year}{2023}\natexlab{}.
\newblock \showarticletitle{{DBREACH}: Stealing from Databases Using Compression Side Channels}. In \bibinfo{booktitle}{\emph{2023 {IEEE} Symposium on Security and Privacy}}. \bibinfo{publisher}{{IEEE} Computer Society Press}, \bibinfo{address}{San Francisco, CA, USA}, \bibinfo{pages}{182--198}.
\newblock
\href{https://doi.org/10.1109/SP46215.2023.10179359}{doi:\nolinkurl{10.1109/SP46215.2023.10179359}}


\bibitem[Huffman(1952)]%
        {huffmancodes}
\bibfield{author}{\bibinfo{person}{David~A. Huffman}.} \bibinfo{year}{1952}\natexlab{}.
\newblock \showarticletitle{A Method for the Construction of Minimum-Redundancy Codes}.
\newblock \bibinfo{journal}{\emph{Proceedings of the IRE}} \bibinfo{volume}{40}, \bibinfo{number}{9} (\bibinfo{year}{1952}), \bibinfo{pages}{1098--1101}.
\newblock
\href{https://doi.org/10.1109/JRPROC.1952.273898}{doi:\nolinkurl{10.1109/JRPROC.1952.273898}}


\bibitem[Igoe and Solinas(2009)]%
        {rfc5647}
\bibfield{author}{\bibinfo{person}{Kevin Igoe} {and} \bibinfo{person}{Jerome Solinas}.} \bibinfo{year}{2009}\natexlab{}.
\newblock \bibinfo{title}{{AES Galois Counter Mode for the Secure Shell Transport Layer Protocol}}.
\newblock \bibinfo{howpublished}{RFC 5647}.
\newblock
\href{https://doi.org/10.17487/RFC5647}{doi:\nolinkurl{10.17487/RFC5647}}


\bibitem[Kelsey(2002)]%
        {FSE:Kelsey02}
\bibfield{author}{\bibinfo{person}{John Kelsey}.} \bibinfo{year}{2002}\natexlab{}.
\newblock \showarticletitle{Compression and Information Leakage of Plaintext}. In \bibinfo{booktitle}{\emph{FSE~2002}} \emph{(\bibinfo{series}{{LNCS}}, Vol.~\bibinfo{volume}{2365})}, \bibfield{editor}{\bibinfo{person}{Joan Daemen} {and} \bibinfo{person}{Vincent Rijmen}} (Eds.). \bibinfo{publisher}{Springer Berlin Heidelberg, Germany}, \bibinfo{address}{Leuven, Belgium}, \bibinfo{pages}{263--276}.
\newblock
\href{https://doi.org/10.1007/3-540-45661-9_21}{doi:\nolinkurl{10.1007/3-540-45661-9_21}}


\bibitem[Krasic et~al\mbox{.}(2022)]%
        {rfc9204}
\bibfield{author}{\bibinfo{person}{Charles~'Buck' Krasic}, \bibinfo{person}{Mike Bishop}, {and} \bibinfo{person}{Alan Frindell}.} \bibinfo{year}{2022}\natexlab{}.
\newblock \bibinfo{title}{{QPACK: Field Compression for HTTP/3}}.
\newblock \bibinfo{howpublished}{RFC 9204}.
\newblock
\href{https://doi.org/10.17487/RFC9204}{doi:\nolinkurl{10.17487/RFC9204}}


\bibitem[Lonvick and Ylonen(2006a)]%
        {rfc4252}
\bibfield{author}{\bibinfo{person}{Chris~M. Lonvick} {and} \bibinfo{person}{Tatu Ylonen}.} \bibinfo{year}{2006}\natexlab{a}.
\newblock \bibinfo{title}{{The Secure Shell (SSH) Authentication Protocol}}.
\newblock \bibinfo{howpublished}{RFC 4252}.
\newblock
\href{https://doi.org/10.17487/RFC4252}{doi:\nolinkurl{10.17487/RFC4252}}


\bibitem[Lonvick and Ylonen(2006b)]%
        {rfc4254}
\bibfield{author}{\bibinfo{person}{Chris~M. Lonvick} {and} \bibinfo{person}{Tatu Ylonen}.} \bibinfo{year}{2006}\natexlab{b}.
\newblock \bibinfo{title}{{The Secure Shell (SSH) Connection Protocol}}.
\newblock \bibinfo{howpublished}{RFC 4254}.
\newblock
\href{https://doi.org/10.17487/RFC4254}{doi:\nolinkurl{10.17487/RFC4254}}


\bibitem[Lonvick and Ylonen(2006c)]%
        {rfc4251}
\bibfield{author}{\bibinfo{person}{Chris~M. Lonvick} {and} \bibinfo{person}{Tatu Ylonen}.} \bibinfo{year}{2006}\natexlab{c}.
\newblock \bibinfo{title}{{The Secure Shell (SSH) Protocol Architecture}}.
\newblock \bibinfo{howpublished}{RFC 4251}.
\newblock
\href{https://doi.org/10.17487/RFC4251}{doi:\nolinkurl{10.17487/RFC4251}}


\bibitem[Lonvick and Ylonen(2006d)]%
        {rfc4253}
\bibfield{author}{\bibinfo{person}{Chris~M. Lonvick} {and} \bibinfo{person}{Tatu Ylonen}.} \bibinfo{year}{2006}\natexlab{d}.
\newblock \bibinfo{title}{{The Secure Shell (SSH) Transport Layer Protocol}}.
\newblock \bibinfo{howpublished}{RFC 4253}.
\newblock
\href{https://doi.org/10.17487/RFC4253}{doi:\nolinkurl{10.17487/RFC4253}}


\bibitem[Miller(2026)]%
        {ietf-sshm-ssh-agent-16}
\bibfield{author}{\bibinfo{person}{Damien Miller}.} \bibinfo{year}{2026}\natexlab{}.
\newblock \bibinfo{booktitle}{\emph{{SSH Agent Protocol}}}.
\newblock \bibinfo{type}{Internet-Draft} draft-ietf-sshm-ssh-agent-16. \bibinfo{institution}{Internet Engineering Task Force}.
\newblock
\urldef\tempurl%
\url{https://datatracker.ietf.org/doc/draft-ietf-sshm-ssh-agent/16/}
\showURL{%
\tempurl}
\newblock
\shownote{Work in Progress}.


\bibitem[Miller et~al\mbox{.}(2026)]%
        {protocolopenssh}
\bibfield{author}{\bibinfo{person}{Damien Miller}, \bibinfo{person}{Markus Friedl}, \bibinfo{person}{Mike Frysinger}, \bibinfo{person}{Todd~C. Miller}, {and} \bibinfo{person}{Darren Tucker}.} \bibinfo{year}{2026}\natexlab{}.
\newblock \bibinfo{title}{This documents {OpenSSH}'s deviations and extensions to the published {SSH} protocol.}
\newblock
\urldef\tempurl%
\url{https://cvsweb.openbsd.org/checkout/src/usr.bin/ssh/PROTOCOL,v?rev=1.60}
\showURL{%
\tempurl}
\newblock
\shownote{Accessed: 2026-04-24}.


\bibitem[Miller et~al\mbox{.}(2025)]%
        {ietf-sshm-chacha20-poly1305-02}
\bibfield{author}{\bibinfo{person}{Damien Miller}, \bibinfo{person}{Simon Tatham}, {and} \bibinfo{person}{Simon Josefsson}.} \bibinfo{year}{2025}\natexlab{}.
\newblock \bibinfo{booktitle}{\emph{{Secure Shell (SSH) authenticated encryption cipher: chacha20-poly1305}}}.
\newblock \bibinfo{type}{Internet-Draft} draft-ietf-sshm-chacha20-poly1305-02. \bibinfo{institution}{Internet Engineering Task Force}.
\newblock
\urldef\tempurl%
\url{https://datatracker.ietf.org/doc/draft-ietf-sshm-chacha20-poly1305/02/}
\showURL{%
\tempurl}
\newblock
\shownote{Work in Progress}.


\bibitem[Miller and Valchev(2007)]%
        {miller-secsh-umac-01}
\bibfield{author}{\bibinfo{person}{Damien Miller} {and} \bibinfo{person}{Peter Valchev}.} \bibinfo{year}{2007}\natexlab{}.
\newblock \bibinfo{booktitle}{\emph{{The use of UMAC in the SSH Transport Layer Protocol}}}.
\newblock \bibinfo{type}{Internet-Draft} draft-miller-secsh-umac-01. \bibinfo{institution}{Internet Engineering Task Force}.
\newblock
\urldef\tempurl%
\url{https://datatracker.ietf.org/doc/draft-miller-secsh-umac/01/}
\showURL{%
\tempurl}
\newblock
\shownote{Work in Progress}.


\bibitem[Nafeez(2018)]%
        {nafeez2018voracle}
\bibfield{author}{\bibinfo{person}{Ahamed Nafeez}.} \bibinfo{year}{2018}\natexlab{}.
\newblock \bibinfo{title}{{Voracle - Compression Oracle Attacks on VPN Tunnels}}.
\newblock \bibinfo{howpublished}{Black Hat USA 2018}.
\newblock
\urldef\tempurl%
\url{https://files.speakerdeck.com/presentations/e71c6e61fada4aa4ab72bce417d33683/Voracle_-_Compression_Oracle_Attacks_on_VPN_Networks.pdf}
\showURL{%
\tempurl}
\newblock
\shownote{Presentation slides, Accessed: 2026-04-30}.


\bibitem[Namprempre et~al\mbox{.}(2006)]%
        {rfc4344}
\bibfield{author}{\bibinfo{person}{Chanathip Namprempre}, \bibinfo{person}{Tadayoshi Kohno}, {and} \bibinfo{person}{Mihir Bellare}.} \bibinfo{year}{2006}\natexlab{}.
\newblock \bibinfo{title}{{The Secure Shell (SSH) Transport Layer Encryption Modes}}.
\newblock \bibinfo{howpublished}{RFC 4344}.
\newblock
\href{https://doi.org/10.17487/RFC4344}{doi:\nolinkurl{10.17487/RFC4344}}


\bibitem[{Oracle}(2026)]%
        {mysql-workbench-ssh-tunnel}
\bibfield{author}{\bibinfo{person}{{Oracle}}.} \bibinfo{year}{2026}\natexlab{}.
\newblock \bibinfo{title}{{MySQL} Workbench Manual---Connect to Server Using SSH Tunneling}.
\newblock
\urldef\tempurl%
\url{https://dev.mysql.com/doc/workbench/en/wb-mysql-connections-methods-ssh.html}
\showURL{%
\tempurl}
\newblock
\shownote{Accessed: 2026-04-24}.


\bibitem[Paterson and Watson(2010)]%
        {EC:PatWat10}
\bibfield{author}{\bibinfo{person}{Kenneth~G. Paterson} {and} \bibinfo{person}{Gaven~J. Watson}.} \bibinfo{year}{2010}\natexlab{}.
\newblock \showarticletitle{Plaintext-Dependent Decryption: A Formal Security Treatment of {SSH}-{CTR}}. In \bibinfo{booktitle}{\emph{EUROCRYPT~2010}} \emph{(\bibinfo{series}{{LNCS}}, Vol.~\bibinfo{volume}{6110})}, \bibfield{editor}{\bibinfo{person}{Henri Gilbert}} (Ed.). \bibinfo{publisher}{Springer Berlin Heidelberg, Germany}, \bibinfo{address}{French Riviera}, \bibinfo{pages}{345--361}.
\newblock
\href{https://doi.org/10.1007/978-3-642-13190-5_18}{doi:\nolinkurl{10.1007/978-3-642-13190-5_18}}


\bibitem[Peon and Ruellan(2015)]%
        {rfc7541}
\bibfield{author}{\bibinfo{person}{Roberto Peon} {and} \bibinfo{person}{Herve Ruellan}.} \bibinfo{year}{2015}\natexlab{}.
\newblock \bibinfo{title}{{HPACK: Header Compression for HTTP/2}}.
\newblock \bibinfo{howpublished}{RFC 7541}.
\newblock
\href{https://doi.org/10.17487/RFC7541}{doi:\nolinkurl{10.17487/RFC7541}}


\bibitem[Pitman(2026)]%
        {awesome-tunneling}
\bibfield{author}{\bibinfo{person}{Anders Pitman}.} \bibinfo{year}{2026}\natexlab{}.
\newblock \bibinfo{title}{awesome-tunneling: List of {ngrok}, {Cloudflare} Tunnel, {Tailscale}, and {ZeroTier} alternatives and other tunneling software and services}.
\newblock
\urldef\tempurl%
\url{https://github.com/anderspitman/awesome-tunneling}
\showURL{%
\tempurl}
\newblock
\shownote{Commit aa10f15, Accessed: 2026-04-24}.


\bibitem[{PostgreSQL Global Development Group}(2026)]%
        {postgresql-ssh-tunnel}
\bibfield{author}{\bibinfo{person}{{PostgreSQL Global Development Group}}.} \bibinfo{year}{2026}\natexlab{}.
\newblock \bibinfo{title}{PostgreSQL 18.3 Documentation---Secure TCP/IP Connections with SSH Tunnels}.
\newblock
\urldef\tempurl%
\url{https://www.postgresql.org/docs/current/ssh-tunnels.html}
\showURL{%
\tempurl}
\newblock
\shownote{Accessed: 2026-04-24}.


\bibitem[Rizzo and Duong(2012)]%
        {CRIME}
\bibfield{author}{\bibinfo{person}{Juliano Rizzo} {and} \bibinfo{person}{Thai Duong}.} \bibinfo{year}{2012}\natexlab{}.
\newblock \bibinfo{title}{The CRIME attack}.
\newblock
\urldef\tempurl%
\url{https://hpc-notes.soton.ac.uk/talks/bullrun/Crime_slides.pdf}
\showURL{%
\tempurl}
\newblock
\shownote{Presentation slides, Accessed: 2026-04-24}.


\bibitem[Ryan et~al\mbox{.}(2023)]%
        {CCS:RHSH23}
\bibfield{author}{\bibinfo{person}{Keegan Ryan}, \bibinfo{person}{Kaiwen He}, \bibinfo{person}{George~Arnold Sullivan}, {and} \bibinfo{person}{Nadia Heninger}.} \bibinfo{year}{2023}\natexlab{}.
\newblock \showarticletitle{Passive {SSH} Key Compromise via Lattices}. In \bibinfo{booktitle}{\emph{ACM CCS 2023}}, \bibfield{editor}{\bibinfo{person}{Weizhi Meng}, \bibinfo{person}{Christian~Damsgaard Jensen}, \bibinfo{person}{Cas Cremers}, {and} \bibinfo{person}{Engin Kirda}} (Eds.). \bibinfo{publisher}{{ACM} Press}, \bibinfo{address}{Copenhagen, Denmark}, \bibinfo{pages}{2886--2900}.
\newblock
\href{https://doi.org/10.1145/3576915.3616629}{doi:\nolinkurl{10.1145/3576915.3616629}}


\bibitem[Schuster et~al\mbox{.}(2017)]%
        {USENIX:SchShmTro17}
\bibfield{author}{\bibinfo{person}{Roei Schuster}, \bibinfo{person}{Vitaly Shmatikov}, {and} \bibinfo{person}{Eran Tromer}.} \bibinfo{year}{2017}\natexlab{}.
\newblock \showarticletitle{Beauty and the Burst: Remote Identification of Encrypted Video Streams}. In \bibinfo{booktitle}{\emph{USENIX Security 2017}}, \bibfield{editor}{\bibinfo{person}{Engin Kirda} {and} \bibinfo{person}{Thomas Ristenpart}} (Eds.). \bibinfo{publisher}{{USENIX} Association}, \bibinfo{address}{Vancouver, BC, Canada}, \bibinfo{pages}{1357--1374}.
\newblock
\urldef\tempurl%
\url{https://www.usenix.org/conference/usenixsecurity17/technical-sessions/presentation/schuster}
\showURL{%
\tempurl}


\bibitem[Song et~al\mbox{.}(2001)]%
        {USENIX:SonWagTia01}
\bibfield{author}{\bibinfo{person}{Dawn~Xiaodong Song}, \bibinfo{person}{David~A. Wagner}, {and} \bibinfo{person}{Xuqing Tian}.} \bibinfo{year}{2001}\natexlab{}.
\newblock \showarticletitle{Timing Analysis of Keystrokes and Timing Attacks on {SSH}}. In \bibinfo{booktitle}{\emph{USENIX Security 2001}}, \bibfield{editor}{\bibinfo{person}{Dan~S. Wallach}} (Ed.). \bibinfo{publisher}{{USENIX} Association}, \bibinfo{address}{Washington, DC, USA}.
\newblock
\urldef\tempurl%
\url{http://www.usenix.org/publications/library/proceedings/sec01/song.html}
\showURL{%
\tempurl}


\bibitem[Tan and Nahata(2013)]%
        {TanNahata2013PETAL}
\bibfield{author}{\bibinfo{person}{Jiaqi Tan} {and} \bibinfo{person}{Jayvardhan Nahata}.} \bibinfo{year}{2013}\natexlab{}.
\newblock \bibinfo{booktitle}{\emph{{PETAL}: Preset Encoding Table Information Leakage}}.
\newblock \bibinfo{type}{{T}echnical {R}eport} CMU-PDL-13-106. \bibinfo{institution}{Carnegie Mellon University Parallel Data Laboratory}.
\newblock
\urldef\tempurl%
\url{https://www.pdl.cmu.edu/PDL-FTP/associated/CMU-PDL-13-106.pdf}
\showURL{%
\tempurl}
\newblock
\shownote{Accessed: 2026-04-24}.


\bibitem[van Goethem et~al\mbox{.}(2020)]%
        {USENIX:GPJV20}
\bibfield{author}{\bibinfo{person}{Tom van Goethem}, \bibinfo{person}{Christina P{\"o}pper}, \bibinfo{person}{Wouter Joosen}, {and} \bibinfo{person}{Mathy Vanhoef}.} \bibinfo{year}{2020}\natexlab{}.
\newblock \showarticletitle{Timeless Timing Attacks: Exploiting Concurrency to Leak Secrets over Remote Connections}. In \bibinfo{booktitle}{\emph{USENIX Security 2020}}, \bibfield{editor}{\bibinfo{person}{Srdjan Capkun} {and} \bibinfo{person}{Franziska Roesner}} (Eds.). \bibinfo{publisher}{{USENIX} Association}, \bibinfo{pages}{1985--2002}.
\newblock
\urldef\tempurl%
\url{https://www.usenix.org/conference/usenixsecurity20/presentation/van-goethem}
\showURL{%
\tempurl}


\bibitem[Vanhoef and Goethem(2016)]%
        {vanhoef2016heist}
\bibfield{author}{\bibinfo{person}{Mathy Vanhoef} {and} \bibinfo{person}{Tom~Van Goethem}.} \bibinfo{year}{2016}\natexlab{}.
\newblock \bibinfo{title}{{HEIST}: {HTTP} Encrypted Information can be Stolen through {TCP}-Windows}.
\newblock \bibinfo{howpublished}{Black Hat USA 2016}.
\newblock
\urldef\tempurl%
\url{https://www.blackhat.com/docs/us-16/materials/us-16-VanGoethem-HEIST-HTTP-Encrypted-Information-Can-Be-Stolen-Through-TCP-Windows-wp.pdf}
\showURL{%
\tempurl}
\newblock
\shownote{White paper, Accessed: 2026-04-24}.


\bibitem[White et~al\mbox{.}(2011)]%
        {SP:WMSM11}
\bibfield{author}{\bibinfo{person}{Andrew~M. White}, \bibinfo{person}{Austin~R. Matthews}, \bibinfo{person}{Kevin~Z. Snow}, {and} \bibinfo{person}{Fabian Monrose}.} \bibinfo{year}{2011}\natexlab{}.
\newblock \showarticletitle{Phonotactic Reconstruction of Encrypted {VoIP} Conversations: Hookt on Fon-iks}. In \bibinfo{booktitle}{\emph{2011 {IEEE} Symposium on Security and Privacy}}. \bibinfo{publisher}{{IEEE} Computer Society Press}, \bibinfo{address}{Berkeley, CA, USA}, \bibinfo{pages}{3--18}.
\newblock
\href{https://doi.org/10.1109/SP.2011.34}{doi:\nolinkurl{10.1109/SP.2011.34}}


\bibitem[White(2021)]%
        {sshtunnel}
\bibfield{author}{\bibinfo{person}{Pahaz White}.} \bibinfo{year}{2021}\natexlab{}.
\newblock \bibinfo{title}{sshtunnel: Pure Python SSH tunnels}.
\newblock
\urldef\tempurl%
\url{https://pypi.org/project/sshtunnel/}
\showURL{%
\tempurl}
\newblock
\shownote{Accessed: 2026-04-24}.


\bibitem[Ziv and Lempel(1977)]%
        {lz77}
\bibfield{author}{\bibinfo{person}{J. Ziv} {and} \bibinfo{person}{A. Lempel}.} \bibinfo{year}{1977}\natexlab{}.
\newblock \showarticletitle{A universal algorithm for sequential data compression}.
\newblock \bibinfo{journal}{\emph{IEEE Transactions on Information Theory}} \bibinfo{volume}{23}, \bibinfo{number}{3} (\bibinfo{year}{1977}), \bibinfo{pages}{337--343}.
\newblock
\href{https://doi.org/10.1109/TIT.1977.1055714}{doi:\nolinkurl{10.1109/TIT.1977.1055714}}


\end{thebibliography}

\iffull
\clearpage
\appendix
\section{Ethical Considerations}
\label{sec:ethics}

\paragraph{Stakeholders} 

We identify several stakeholders affected by our work. Most directly affected are SSH users---operators, developers, and administrators who rely on SSH for remote shells, port forwarding, and automation---whose passwords or other secrets can be recovered when our attack preconditions hold. Closely related are the operators of services reached through SSH tunnels since secrets can leak through our attack even when the tunneled service is itself secure. SSH implementation vendors and maintainers are stakeholders because they decide which features ship and which defaults apply, both of which influence exploitability. On the standardization side, the IETF sshm working group and the authors of RFCs 4251--4254 are stakeholders because the root cause is structural to the SSH protocol. Within the research community, we identify the authors of prior compression side-channel attacks (in particular, CRIME and BREACH), ourselves, and the broader security research community. Potential adversaries are stakeholders as well, since publication and the accompanying artifact allow mounting the attack. Finally, the public is an indirect stakeholder: successful recovery of secrets such as database passwords can lead to subsequent breaches affecting people whose data is stored in those systems.

\balance

\paragraph{Impacts} 

For SSH users, our work presents a new way for attackers to recover secrets in standards-compliant configurations. Before publication this risk exists but is hidden; after publication it is publicly known, which allows vendors to ship fixes and operators to deploy countermeasures. Operators of tunneled services are affected indirectly: their secrets can leak even when their service is otherwise secure, which is unlikely to be part of their threat model. SSH vendors and maintainers will need to update documentation, default settings, and, in some cases, code, which is additional work in the short term but benefits the security of their implementations in the long term. For the IETF sshm working group and the authors of RFCs~4251--4254, our findings show that the Connection Protocol and the Binary Packet Protocol cannot be reasoned about separately, which may start a discussion at the standards level. The wider research community, including the authors of CRIME and BREACH, gains novel insight into SSH channel multiplexing as a security boundary that had not been studied before. For us as authors, the impact is the responsibility that comes with publishing this work, in particular performing a coordinated vulnerability disclosure. Potential adversaries may benefit from the published artifacts until countermeasures are deployed in ways that harm other stakeholders. The public is affected indirectly but most seriously: if secrets such as database passwords are recovered, the resulting breaches can expose the personal data of people.

\paragraph{Mitigations} 

To limit harm, we took several steps. All experiments were carried out against infrastructure under our control. No production system, third-party server, or non-consenting user was attacked, probed, or observed, and no user data was collected. For disclosure, we first directly contact the maintainers of those implementations in \cref{tab:implementations} whose defaults make their users vulnerable---in particular the clients that prefer compression or bind publicly by default---and share details about our findings and potential countermeasures discussed in \cref{sec:countermeasures}. Second, after these vendors have had a reasonable mitigation window, we make a public disclosure on the IETF sshm mailing list. This second step aims to raise awareness across the broader SSH community, possibly starting a discussion of a standards-level response, analogous to the TLS community's removal of compression after CRIME. We acknowledge that the artifacts still carry some misuse risk; we limit this by stating prominently that users can mitigate the attack immediately by disabling SSH compression on either end of the connection.

\paragraph{Decision to Publish} 

We decided that the publication of this work, alongside the mitigations described above, is ethically justified. The root cause has been part of SSH since channel multiplexing and compression were first specified, and CRIME and BREACH show that such gaps are eventually rediscovered. Not publishing would leave the users at risk, which we deem to be the worst outcome.

\section{Open Science}
\label{sec:openscience}

We are committed to open science, allowing for independent reproduction of our findings. As such, we make our artifacts available under the Apache\nobreakdash-2.0 license, fostering further research in the area of SSH and network security in general. Our artifacts include:

\begin{itemize}
    \item A unified attack engine implemented in Python that performs the generic attack described in \cref{sec:attack}.
    \item Adapter classes to integrate the unified attack engine with the three scenarios described in \cref{sec:evaluation}.
    \item Docker images and compose files for setting up the evaluation environment.
    \item Integration tests that recover a single password to verify algorithmic correctness, additional helper scripts, and documentation.
    \item Our evaluation artifacts, in particular the raw statistics from which we compiled \cref{tab:evaluation}.
\end{itemize}
Available here: \urlArtifacts
\fi

\end{document}